\documentclass[aps,prb,twocolumn,groupedaddress,10pt]{revtex4-2}
\usepackage{graphicx}
\usepackage{multirow}
\usepackage{amssymb}

\begin{document}


\title{Isostructural phase transition and equation of state of type-I and type-VIII metallic sodium borosilicide clathrates}


\author{M. Demoucron$^{1*}$}
\author{S. Pandolfi$^{1}$}
\author{ Y. Guarnelli$^{1}$}
\author{B. Baptiste$^{1}$}
\author{P. Chauvigné$^{2}$}
\author{N. Guignot$^{2}$}
\author{D. Portehault$^{3}$}
\author{T. A. Strobel$^{4}$}
\author{T.B. Shiell$^{4}$}
\author{M. Bykov$^{4}$}
\author{W.A. Crichton$^5$}
\author{Y. Le Godec$^{1*}$}
\author{A. Courac$^{1}$}

\email[]{martin.demoucron@sorbonne-universite.fr, yann.le\_godec@sorbonne-universite.fr, alexandre.courac@sorbonne-universite.fr}
\affiliation{$^{1}$ Sorbonne Université, MNHN, UMR CNRS 7590, IMPMC, 75005 Paris, France.}
\affiliation{$^{2}$Synchrotron SOLEIL, 91192 Saint Aubin, France.}
\affiliation{$^{3}$LCMCP, Sorbonne Université, UMR CNRS 7574, 75005 Paris, France.}
\affiliation{$^{4}$Earth and Planet Laboratory, Carnegie Institution for Science, Washington, DC 20015, USA.}
\affiliation{$^{5}$ESRF, The European Synchrotron, Grenoble, France.}


\date{\today}

\begin{abstract}

Electronic properties of silicon-based clathrates can be tuned by boron incorporation into the silicon cage network. Sodium borosilicides clathrate outstands with uncommon stoichiometry and expected metallic properties, in contrast to other alkali metal semiconductive Zintl borosilicides. In this study, we report an experimental investigation of the high-pressure behavior of type-I and type-VIII sodium borosilicide clathrates. An isostructural phase transition, marked by an abrupt volume collapse at 13 GPa, is observed exclusively in type-I sodium borosilicide clathrates. This transition is attributed to the pressure-induced diffusion of silicon atoms from the Si(6\textit{c}) site. This mechanism provides the first experimental validation of a transition predicted theoretically for this class of materials. Isostructural phase transitions were only observed in type-I borosilicide. In contrast, the type-VIII borosilicide phase exhibits conventional elastic compression. The metallic character was established using reflectance spectroscopy over a wide energy range, in good agreement with crystallographic data on the boron content.

\end{abstract}


\maketitle



Silicon intermetallic clathrates are cage-based framework materials that hold great promise for applications in photovoltaics \cite{himeno_optical_2013, he_si-based_2014}, superconductivity \cite{kawaji_superconductivity_1995}, batteries \cite{dopilka_structural_2021} and gas storage \cite{neiner_synthesis_2010}. 
For a relatively small number of zeolite-type structure, a wide range of chemical compositions is possible. Most reported clathrates crystallize in the cubic type-I structure ($P\text{m}\overline{3}\text{n}$ space group), which consist of an arrangement of two small 5$^{12}$ cages (dodecahedra, 12 pentagonal faces) and six larger 5$^{12}$6$^2$ cages (tetrakaidekahedra, 12 pentagonal faces and 2 hexagonal faces) \cite{dolyniuk_clathrate_2016}. On the other hand, only a few compounds with a type-VIII structure ($I\overline{4}\text{3}m$ space group) exist. It features an alternative polyhedral arrangement in a body-centered cubic structure made of uncommon asymmetric cages with 20+3 vertices/atoms \cite{shevelkov_zintl_2011}. Due to the asymmetry of type-VIII clathrate cages, small cavities surrounded by eight atoms are also part of the structure \cite{leoni_modelling_2003}, thus type-VIII is not a true polyhedral clathrate in sense of the cages fully tiling three-dimensional space. In both structure types, the cages are occupied by electropositive guest atoms that donate their electrons to the covalent silicon-based framework.

\begin{figure*}
    \centering
    \includegraphics[width=1\linewidth]{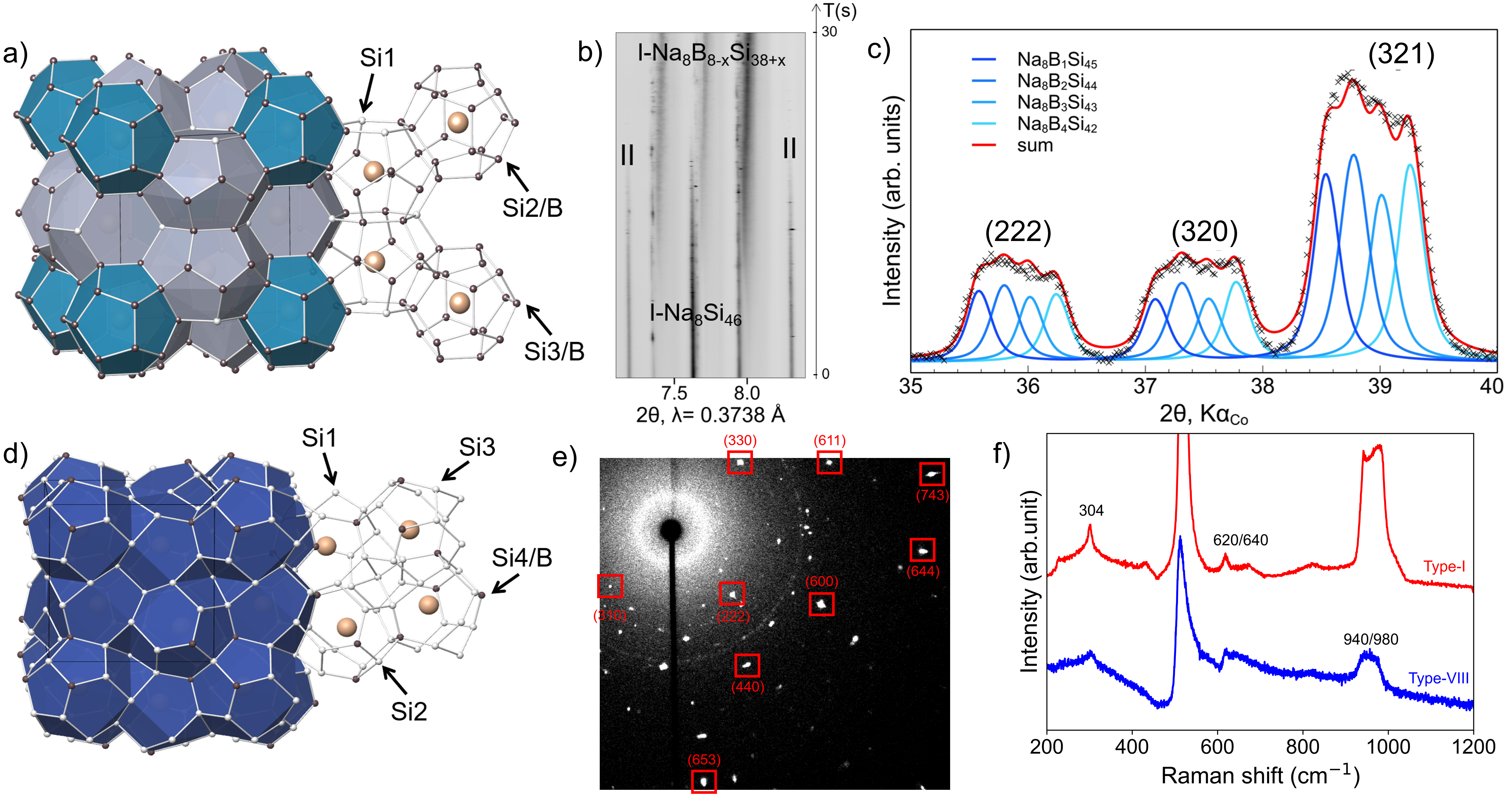}
    \caption{a) Crystal structure of type-I Na$_8$B$_x$Si$_{46-x}$ clathrate. b) Evolution of the \textit{in situ} XRD diffraction signal (between 7.1 and 8.4°) through high-pressure high temperature conditions (large volume press experiment, 1100~K – 3.5~GPa). c) XRD powder pattern  (between 35 and 40°) of the sample synthetized at 3.5~GPa and 1150~K, the intensities of the computed phases are calculated from Na$_8$B$_1$Si$_{45}$ to Na$_8$B$_4$Si$_{42}$ fixed stoichiometry with boron atom inside the 16\textit{i} and 24\textit{k} Wyckoff sites.  d) Crystal structure of type-VIII Na$_8$B$_x$Si$_{46-x}$ clathrate. e) Part of type-VIII Na$_8$B$_{4.1}$Si$_{41.9}$ single-crystal XRD detector image with some indexed reflections. f) Raman spectrum at ambient condition of type-I borosilicide powder sample and Na$_8$B$_{4.1}$Si$_{41.9}$ type-VIII single-crystal clathrate.}
    \label{DRX computed NaBSi}
\end{figure*}

Only a few type-VIII silicon-based clathrates are known, and no binary M$_8$Si$_{46}$ (M = Na, K, or Ba) compounds having the type-VIII structure have ever been reported. All known type-VIII clathrates require the incorporation of additional group-13 elements such as Sr$_8$Al$_x$Ga$_{16-x}$Si$_{30}$ \cite{kishimoto_synthesis_2008},   
Sr$_8$Al$_x$Ga$_y$Ge$_{46-x-y}$ \cite{sasaki_synthesis_2009} and Ba$_8$Ga$_{16}$Sn$_{30}$ \cite{suekuni_simultaneous_2008}. Also, compounds with approximately identical stoichiometry that crystallize in both type-I and type-VIII frameworks are rare; a notable example is the borosilicide Na$_8$B$_4$Si$_{42}$ clathrate, which has been reported in both structural types \cite{hubner_borosilicide_2022}. In these clathrates, alkali metals donate electrons to compensate for the octet-rule deficiency induced by boron substitution for silicon inside the four-bonded atom framework. In the type-I model structure, boron atoms substitute silicon within the 16\textit{i} and 24\textit{k} Wyckoff sites (See Figure \ref{DRX computed NaBSi}.a). In the type-VIII structure, boron atoms substitute silicon at the 8\textit{i} site (See Figure \ref{DRX computed NaBSi}.d). 
In this work, we studied a series of ternary Na–B–Si compounds under high-pressure, high-temperature conditions, including type-I and type-VII borosilicides clathrates. We investigated their behavior under high quasi-hydrostatic pressure and revealed an isostructural phase transition driven by the pressure-induced diffusion of silicon and creation and vacancies. 

With one exception \cite{jung_k7_2007}, the synthesis of boron-doped silicon clathrates requires the application of high-pressure, as demonstrated for Rb$_8$B$_8$Si$_{38}$ \cite{hubner_cage_2021}, Cs$_8$B$_8$Si$_{38}$ \cite{hubner_mastering_2021} and Na$_8$B$_4$Si$_{42}$ \cite{hubner_borosilicide_2022}. 
Compression promotes inter-diffusion and partial substitution of silicon by boron, thereby tuning the band-gap energy and strengthening the covalent intermetallic framework. Consequently, our borosilicide clathrates samples were synthesized using large volume multi-anvil presses in order to reach moderate pressure below 4 GPa. Type-I Na$_8$B$_x$Si$_{46-x}$ samples were obtained at 3.5~GPa and 1150~K, whereas the type-VIII Na$_8$B$_{4.1(1)}$Si$_{41.9(1)}$ sample was produced at higher pressure and temperature (4~GPa and 1500~K). Finally, a boron-free type-I Na$_8$Si$_{46}$ clathrate was synthesized  via a conventional route as a reference sample \cite{kasper_clathrate_1965, song_straightforward_2021}. All the details of these procedures are described in the Supplementary Material \cite{demoucron_isostructural_2025}. The boron content of all synthesized samples and the associated bulk moduli and pressure derivatives discussed below are summarized in Table \ref{table NaBSi}. 

\begin{table*}
    \centering
    \begin{tabular}{ccccccc}
    \hline
        \multirow{2}{*}{\textbf{Composition}} & \multirow{2}{*}{\textbf{Synthesis conditions}} & \multirow{2}{*}{\textbf{a$_0$(Å)}} & \multicolumn{3}{c}{\textbf{Equation of state}} & \multirow{2}{*}{\textbf{Reference
        }}  \\ \cline{4-6}
        ~ & ~ & ~ & P$_{\text{range}}$ \!(GPa) & B$_0$ \!(GPa) & B$_0$$^{\prime}$ & ~   \\ \hline
        Na$_8$Si$_{46}$ - type-I & See text & 10.200(1)* & [0-20] & 64.7(18) & 6.8(5) & This work   \\ \hline
        Na$_8$B$_{4.8}$Si$_{41.2}$ - type-I &  & 9.936(14) & [0-13] & 86.6(17) & 2.5(3) & This work   \\
        Na$_8$B$_{3.7}$Si$_{42.3}$ - type-I & 3.5 \!GPa/1150 \!K & 9.999(5) & [0-13] & 99(5) & 1.5(8) & This work   \\
        Na$_8$B$_{2.9}$Si$_{43.1}$ - type-I &  & 10.041(2) & [0-12] & 88(11) & 3.7(18) & This work   \\ \hline
        Na$_8$B$_{4.1(1)}$Si$_{41.9(1)}$ - type-VIII & 4 \!GPa/1500 \!K & 9.699(1)* & [0-22] & 90.2(13) & 3.8(2) & This work   \\ \hline
        Na$_8$B$_{4.1(7)}$Si$_{41.9(7)}$ - type-I & 5 \!GPa/1200 \!K & 9.977(1) & - & - & - &  \cite{hubner_borosilicide_2022}    \\ 
        Na$_8$B$_{4.2(1)}$Si$_{41.8(1)}$ - type-VIII & 6 \!GPa/1200 \!K & 9.7187(2) & - & - & - &  \cite{hubner_borosilicide_2022}   \\ \hline
    \end{tabular}
    \caption{Sample properties and P-V equation of state data (2$^{nd}$ order \textit{Vinet} EoS was used \cite{vinet_compressibility_1987}). *:  fixed a$_0$ from powder XRD at ambient conditions.}
    \label{table NaBSi}
\end{table*}


The \textit{in situ} high-pressure, high-temperature XRD measurements at 1100~K (See Figure \ref{DRX computed NaBSi}.b and see Supplementary Material for starting material \cite{demoucron_isostructural_2025}) reveal the emergence of spots corresponding to individual type-I clathrates with reduced lattice parameters. Type-I borosilicide clathrates formed at the expense of the boron-free Na$_8$Si$_{46}$ (type-I) and Na$_{24}$Si$_{136}$ (type-II) clathrates as respectively identified in references \cite{kurakevych_na-si_2013} and \cite{yamanaka_high-pressure_2014}. The associated \textit{ex situ} XRD of the recovered sample is represented on Figure \ref{DRX computed NaBSi}.c and reveals a wide range of boron content within type-I Na$_{8}$B$_x$Si$_{46-x}$ clathrates at ambient conditions. The broadening of powder XRD reflections indicates a mixture of compounds or solid solutions with various boron concentrations, rather than a non-homogeneous boron distribution within one phase, with local regions exhibiting lattice parameters ranging from 10.147 down to 9.973~Å—in contrast with the Si-Ge system and other solid-solutions \cite{gerin_structural_2023}. 
The significant volume reduction compared to the boron-free type-I Na$_8$Si$_{46}$ clathrate (a$_0$~\!=~\!10.2004(3)~Å) provides unambiguous evidence of the presence of boron within the structure \cite{zhang_covalent_2018} as it was observed under extreme conditions. Le Bail refinement of the \textit{ex situ} XRD pattern was performed using four phases, with boron atoms occupying the 16\textit{i} and 24\textit{k} Wyckoff sites, as reported by \textit{Hübner et al.} \cite{hubner_borosilicide_2022}. By default, four phases were used to describe the XRD signal, but the number of phases within this sample is unknown and could encompass more components. Nonetheless, at high 2$\theta$ angles, where many peaks overlap, the calculated Na$_8$B$_x$Si$_{46-x}$ phases accurately reproduce the XRD signal (see Supplementary Material, Figure S2 \cite{demoucron_isostructural_2025}), thereby confirming the validity of the fit. Nonhomogeneous boron distribution was already suggested by \textit{Jung et al.} \cite{jung_k7_2007, jung_impact_2021} in the K-B-Si system. Distinct phases with different boron contents (K$_{8-x}$B$_y$Si$_{46-y}$) describe the same continuous range of boron content within the type-I clathrate. Each measured composition (from K$_{6.80(2)}$B$_{6.4(5)}$Si$_{39.6(5)}$ to K$_{7.85(2)}$B$_{7.8(1)}$Si$_{38.2(1)}$) corresponds to a distinct lattice parameter. 

On the other hand, synthesis under higher pressure and temperature conditions stabilizes the unique type-VIII borosilicide phase, along with two other distinct type-I borosilicide clathrates. Type-VIII borosilicide clathrate crystals were extracted from powder to conduct single-crystal XRD under ambient conditions. Single-crystal XRD signal (See Figure \ref{DRX computed NaBSi}.e) clearly confirms a structure and a composition similar to that previously reported \cite{hubner_borosilicide_2022}. Raman spectroscopy at ambient conditions (See Figure \ref{DRX computed NaBSi}.f) was also performed on the same crystal. A similar Fano profile and anti-resonance dip from low-energy side is observed in heavily doped p-type silicon around 512 cm$^{-1}$\cite{kumar_effect_2021}. This asymmetric line shape appears exclusively in the type-VIII clathrate, in agreement with a metallic character \cite{rani_fano-type_2024}. The band near 620/640~cm$^{-1}$ serves as a fingerprint of boron incorporation in the silicon network \cite{kumar_effect_2021}. High-temperature Raman spectroscopy confirmed the thermal stability of this single-crystal up 1273~K (see Supplementary Material, Figure S7 \cite{demoucron_isostructural_2025}), while Fano feature is clearly observed up to 873~K. FTIR spectroscopy was futher performed on the same Na$_8$B$_{4.1}$Si$_{41.9}$ type-VIII single-crystal in both the near- (500-7000~cm$^{-1}$) and mid-IR (8500-22500~cm$^{-1}$) ranges (See Supplementary Material, Figure S8 \cite{demoucron_isostructural_2025}). The reflectance spectra display a metallic response, similar to that of gold or aluminum, with comparable trend \cite{shanks_optics_2016}. No transmission is observed in either spectral range, consistent with metallic behavior, in contrast with narrow bandgap like BC8 silicon \cite{zhang_bc8_2017}.

The Zintl-Klemm concept \cite{nesper_zintl-klemm_2014} would predict a Na$_x$B$_x$Si$_{46-x}$ composition for both clathrate types, but according to previously reported stoichiometry \cite{hubner_borosilicide_2022} and our type-VIII single-crystal XRD measurement (discussed later), the Na$_8$B$_x$Si$_{46-x}$ stoichiometry is retained for all samples. The boron concentration in each clathrate cannot be determined by our powder XRD data, therefore, we assume that the sodium borosilicide crystals obey \textit{Vegard’s} law \cite{vegard_konstitution_1921}, previously validated for K$_{8-x}$B$_y$Si$_{46-y}$ type-I clathrates \cite{jung_impact_2021}, which describes a linear relationship between the lattice parameter and the boron content. Using the previously reported sodium borosilicide lattice parameter \cite{hubner_borosilicide_2022}, we evaluate the boron contents of the three type-I borosilicide clathrates identified in the sample as 2.9, 3.7 and 3.8 at.\% (See Supplementary Material, Figure S3 \cite{demoucron_isostructural_2025}). Type-VIII borosilicide clathrate single crystals could be extracted from the crushed sample synthetized at 4~GPa and 1500~K and subsequent single crystal XRD analyses were performed. The refined stoichiometries and the lattice parameters are found to be very close to the structure reported by \textit{Hübner et al.} \cite{hubner_borosilicide_2022}. Boron atoms substitute silicon only in the 8\textit{c} Wyckoff site and no residual boron was found into other sites. Furthermore, a significant electron density has been identified at the center of the cages. From single crystal XRD the number of sodium atoms is fixed at one per cage, and both the Na$_8$B$_{4.1(2)}$Si$_{41.9(2)}$ and Na$_8$B$_{4.51(10)}$Si$_{41.49(10)}$ stoichiometries were refined on two different samples (See Supplementary Material Tab. S1 - Tab. S3 \cite{demoucron_isostructural_2025}).

The behavior under pressure of type-I and type-VIII borosilicide clathrates was studied in order to probe their previously unknown elastic properties and reveal possible isostructural phase transitions. The samples were loaded into diamond anvil cells (DACs) equipped with 400~µm culet diamonds to reach 20~GPa. 40~µm thick rhenium gaskets with a 150~µm diameter hole were used and filled with neon or argon gas as pressure-transmitting media. The increase of pressure was achieved by a gas membrane. A small amount of gold powder was placed as a pressure calibrant inside each DAC. \textit{In situ} XRD analyses were carried out at synchrotron SOLEIL on the PSICHÉ beamline with a beam wavelength of 0.3738~\!Å. The beam size was approximately 10~\!$\times$~\!10~µm, while the sample measured around 50~\!µm~\!$\times$~\!50~\!µm~\!$\times$~\!20~\!µm. This \textit{in situ} characterization enabled us to follow the lattice evolution and determine the equation of state (EoS) of compounds from precise Rietveld refinements with 0.01° 2$\theta$ resolution (R-factor range from 1 to 5, see Supplementary Material Fig S5 \cite{demoucron_isostructural_2025}). The boron-free Na$_8$Si$_{46}$ clathrate exhibited a powdered diffraction signal, whereas the type-I and type-VIII borosilicide clathrates produced spotty diffraction patterns. Nevertheless, this was sufficient for diffractogram integration. To obtain accurate pressure values, XRD patterns of gold were acquired before and after measuring sample, and the average gold cell volume was used as a pressure gauge. The Anderson EoS of gold \cite{anderson_anharmonicity_1989} was employed to determine the pressure at each experimental point. Rietveld refinements were performed using the MAUD software \cite{lutterotti_maud_2025} to extract both sample and pressure calibrants lattice parameters. Additional pressure calibrants with known EoS, including cubic silicon \cite{anzellini_quasi-hydrostatic_2019}, solid neon \cite{dewaele_high_2008}, and solid argon \cite{dewaele_stability_2021} were also used.

Synchrotron XRD patterns were recorded over a pressure of $\sim$0.5~GPa to 20~GPa at 300~K. The initial XRD patterns of the compounds, collected from different regions of the cell, are shown in Figure S4 of the Supplementary Material \cite{demoucron_isostructural_2025}. The presence of elemental cubic silicon-I inside the sample synthetized at 3.5~GPa and 1150~K, likely results from incomplete reaction during clathrate formation. 
When silicon was detected, the cubic-to-hexagonal phase transition (Si-I → Si-V) occurred at 13 GPa, while the intermediate tetragonal and orthorhombic phases (Si-II and Si-XI) were not always observed. A trace amount of rhenium, used as a DAC gasket, is also discernible through the observation of minor and broad peaks situated around Q~\!=~\!2.9~Å$^{-1}$. The small beam size limited the number of grains probed, resulting in texture on the 2D image detector and small deviations of peak intensities from theoretical ratios. However, the presence of type-I and type-VIII clathrate phases was clearly identified. 

\begin{figure*}
    \centering
    \includegraphics[width=1\linewidth]{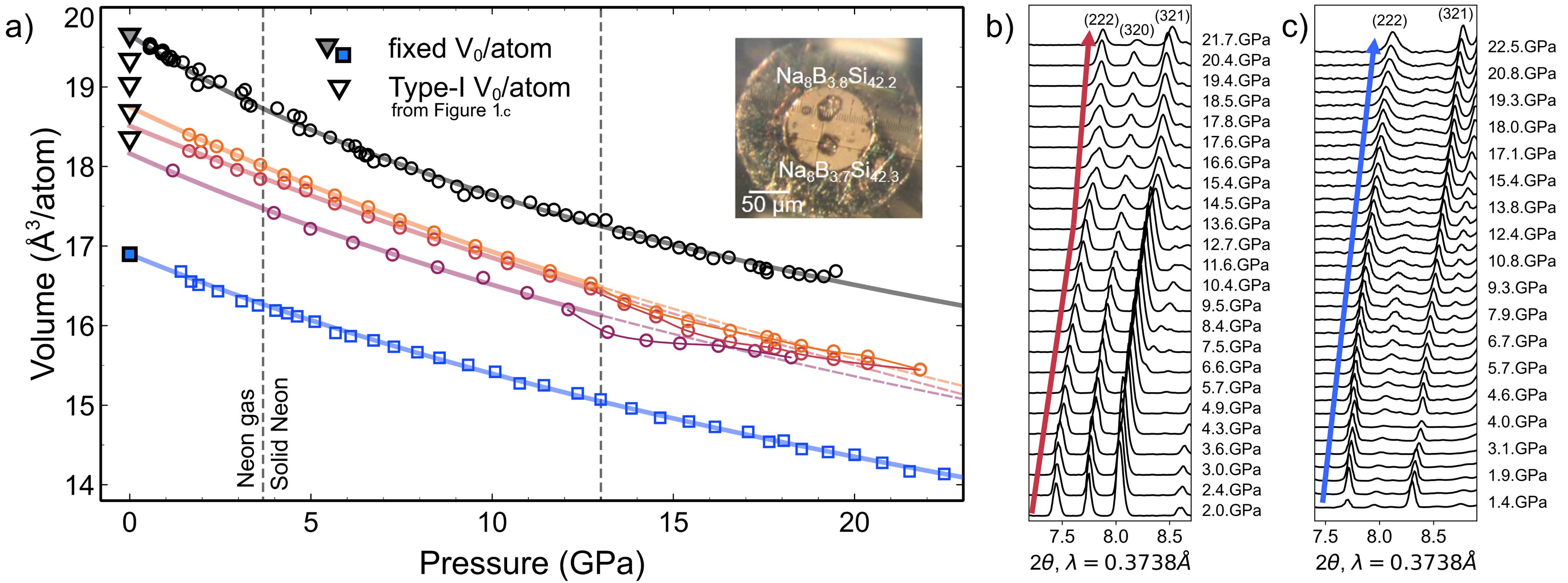}
    \caption{a) The P-V experimental data and their corresponding Vinet EoS fitting along with the picture of Na$_8$B$_{4.8}$Si$_{41.2}$ and Na$_8$B$_{3.7}$Si$_{42.3}$ samples inside the DAC. The black curve corresponds to the boron-free type-I Na$_8$Si$_{46}$ clathrate. The blue curve corresponds to the type-VIII Na$_8$B$_{4.1(1)}$Si$_{41.9(1)}$ clathrate. All the other samples correspond to the Na$_8$B$_x$Si$_{46-x}$ type-I clathrates ranging from orange to purple according to their initial lattice parameter. The initial V$_0$ volume of the borosilicides type-I clathrates synthetized at 3.5~GPa and 1150~K are unknown and, thus, cannot be fixed during the equations of state fitting. The pressure and volume/atom error bars are included in the width of markers. For type-I borosilicide clathrates, above 13 GPa, the solid lines serve as guides to highlight the volume collapse. b) Waterfall plotting of Na$_8$B$_{3.7}$Si$_{42.3}$ type-I raw data (222), (320) and (322) XRD triplet. c) Waterfall plotting of Na$_8$B$_{4.2}$Si$_{41.8}$ type-VIII raw data (222) and (321) XRD doublet.}
    \label{EoS NaBSi}
\end{figure*}

Figure \ref{EoS NaBSi}.a shows the pressure-volume (P-V) experimental data for all clathrate samples up to 22~GPa, along with a picture of one DAC loading.
In three different static compression experiments, four distinct type-I clathrate phases, recovered from high-pressure, were investigated: one sodium silicide and three Na–B–Si phases. All phases exhibit simultaneous volume reduction without any crossover between their P–V curves.
The data are continuous up to 13~GPa and fit well to the Vinet EoS. Above 13~GPa, only Na$_8$Si$_{46}$ exhibits a continuous compression behavior, whereas the three type-I borosilicide clathrates phases display a sudden volume collapse near 13~GPa. This phenomenon was reproducible in two different DAC experiments.
Neon, used as a pressure-transmitting medium, solidifies above 4 GPa and thereafter serves as a pressure gauge. The Si-I phase is observed below 13 GPa, prior to the Si-I → Si-V phase transition. Those pressure gauges, Si \cite{anzellini_quasi-hydrostatic_2019}, Ne \cite{dewaele_high_2008} and Ar \cite{dewaele_stability_2021} are essential to testify the possible stress over the cell arising during transformations by comparison to the gold gauge.
They confirm that the observed volume collapse is not due to any pressure calibration error. The extracted pressure values from all calibrants are compared with those obtained from gold (See Supplementary Material, Figure S6 \cite{demoucron_isostructural_2025}). The possibility of a pressure anomaly caused by the large volume reduction during the Si-I to Si-V phase transition was discarded, due to the significant change in the intensity of borosilicide clathrate peaks, indicating a structural rearrangement. Also, any discontinuity between the measured pressure inside the cell and the pressure of the gas DAC membrane was observed. 

\begin{figure*}
    \centering
    \includegraphics[width=0.75\linewidth]{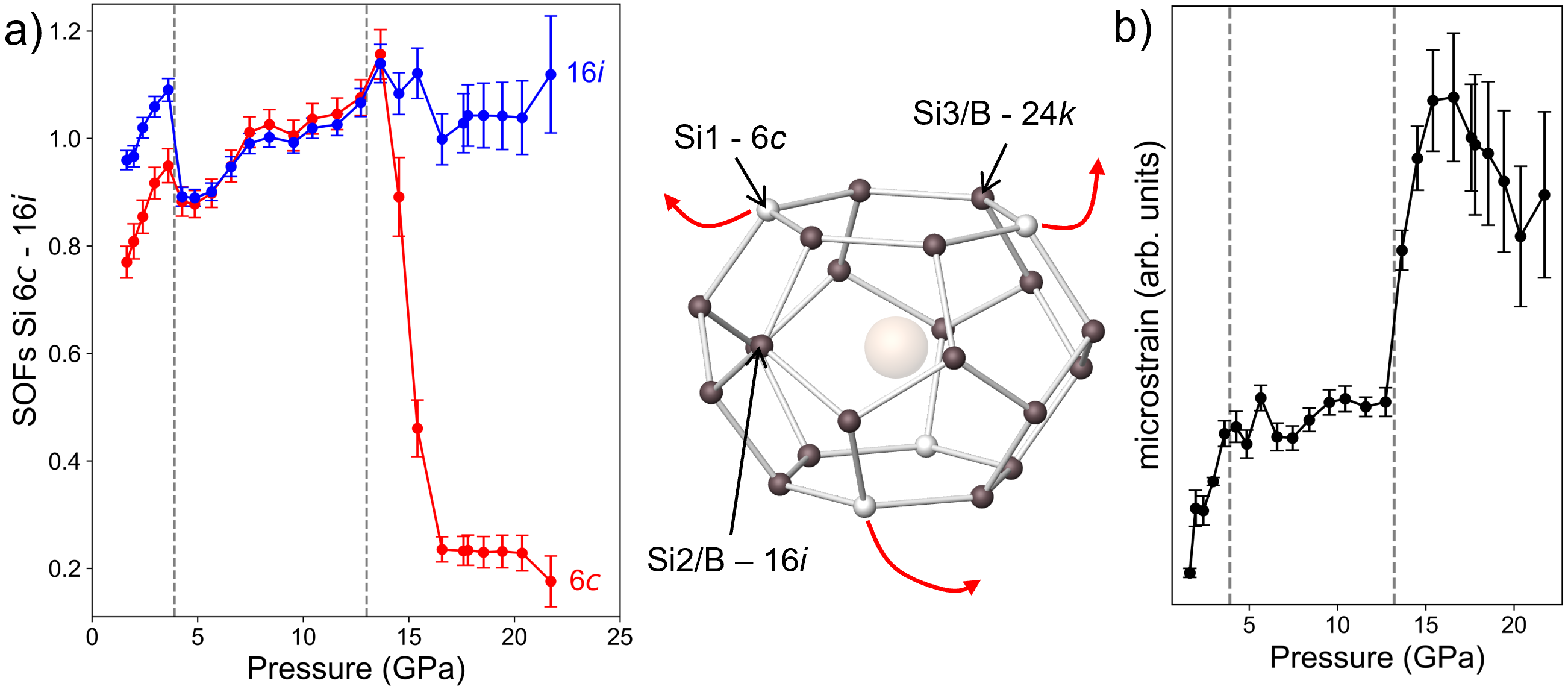}
        \caption{a) Site-occupancy factors for the 6d and 16i Wyckoff sites with a schematic representation of Si diffusion into Na$_8$B$_{4.8}$Si$_{41.2}$ sample and b) induced microstrain calculated from peaks enlargement.}
    \label{strain NaBSi}
\end{figure*}

The volume collapse at 13~GPa is observed without any change in crystal symmetry (See Figure \ref{EoS NaBSi}.b). 
Analysis of the XRD patterns during compression shows that the shape and position of each diffraction spot remain nearly unchanged across the transition, even after the volume collapse. The volume collapse is therefore attributed to an isostructural phase transition, similar to those previously observed for in K-Si and Ba-Si type-I clathrates \cite{debord_isostructural_2021, san-miguel_pressure_2002, miguel_pressure-induced_2005, toulemonde_high_2011}. The theoretical study of \textit{Iitaka} \cite{iitaka_pressure-induced_2007} suggests a possible mechanism of this isostructural phase transition, which starts with the creation of Si vacancies under high-pressure, especially at the 6\textit{c} silicon site. In this mechanism, free silicon atoms can diffuse and form silicon-rich phases such as B-doped diamond-like silicon, at the grain boundaries. Because only a limited number of $hkl$ reflections  ($\leq$~8) were available for refinement, only two site occupancy factors (SOFs) could be refined simultaneously to ensure reliable and meaningful results. Figure \ref{strain NaBSi}.a shows the refined SOFs for silicon (16\textit{i}) and (6\textit{c}) sites. Some SOFs values above unity can be attributed by spotty diffraction patterns and the contribution of gas diffuse signal before 4~GPa, leading to an incorrect evaluation of peak intensities. However, the trends of SOFs after 13~GPa show clear transition, especially the decrease of occupancy for the Si(6\textit{c}) site in the tetrahedral framework. The others SOFs for the Si(16\textit{i}) site exhibit values centred around one, confirming that the silicon vacancies are coming from a preferred site. This reduction in atomic density leads to the corresponding volume collapse, which, however, appears to be compensated at higher pressures (See Figure \ref{EoS NaBSi}.a). This phenomenon has never been observed in any Si-based clathrates and could be the consequence of a higher boron content per cage after silicon diffusion, thereby rendering the framework more covalent.

This isostructural transition induces microstrain within the sample due to cage rearrangement (See Figure \ref{strain NaBSi}.b). At low pressure, prior to Ne solidification, both the SOFs and microstrain exhibit anomalous behavior, likely arising from diffuse scattering of the gas pressure-transmitting medium.
Overall, the evolution of cell volume, site occupancies, and microstrain supports the mechanism proposed by \textit{Itaka} \cite{iitaka_pressure-induced_2007}, in which silicon atoms migrate from the tetrahedral covalent framework at Si(6\textit{c}) site. The initial SOF error bars do not take into account the fact that the spotty diffraction signal can influence the peak intensity distribution, which may  fluctuate from the theoretical values. However, at the isostructural phase transition, all diffraction spots are found to be identical and at the same position. The trends in their intensities, and thus the SOFs, can therefore be reliably analyzed. The observed sudden change of compressibility could be thus explained by the formation of Si/B vacancies within the cage framework.
The same isostructural phase transition phenomena was not observed in the boron-free Na$_8$Si$_{46}$ type-I clathrate and within the Na$_8$B$_{4.1(1)}$Si$_{41.9(1)}$ type-VIII clathrate (See Figure\ref{EoS NaBSi}.c). This can be illustrated by a lack of defects or vacancies within these two compounds, explained by different route syntheses. Also, it is not excluded that similar isostructural phase transitions might be observed at pressure above 20~GPa.

The values of bulk moduli of three observed borosilicides type-I clathrate phases were then calculated from the P-V curves at pressures below the isostructural phase transition, i.e. for the 0-13~GPa range. All data were fitted using the third-order Vinet equation of state implemented in the EoSFit7 software \cite{gonzalez-platas_eosfit7-gui_2016}. The ambient-pressure lattice parameters, fitting pressure ranges, and corresponding compressibility parameters are summarized in Table \ref{table NaBSi}. For samples with unknown ambient lattice parameters (a$_0$), these values were left free during fitting. Boron incorporation into the type-I framework increases the bulk modulus. This phenomenon can be attributed to the strong covalent bond between boron and silicon. On the other hand, the type-VIII Na$_8$B$_{4.1}$Si$_{41.9}$ clathrate exhibits a lower lattice parameter in comparison to all type-I clathrates, and 10.7~\!\% higher crystallographic density than the Na$_8$Si$_{46}$ type-I clathrate. Despite this, its bulk modulus is comparable to that of type-I clathrates. The borosilicide clathrates exhibit B$_0$ values of bulk moduli that fall within the range of \textit{d}-Si \cite{anzellini_quasi-hydrostatic_2019, mcmahon_pressure_1994}. This fact further corroborates the notion that isolated grains of type-I phases are characterised by distinct properties, thus underscoring the necessity for dedicated research in this area.

Finally, we presented distinct syntheses of boron-free and borosilicide type-I and type-VIII clathrates. 
Time-resolved \textit{in situ} XRD enabled us to identify at least three different  type-I borosilicide clathrates, which exhibit an abrupt volume collapse above 13 GPa. We attribute this behavior to an isostructural transition driven by atomic migration from the tetrahedral. In contrast,  type-I Na$_8$Si$_{46}$ and type-VIII Na$_8$B$_{4.1}$Si$_{41.9}$ clathrates remain structurally stable up to 20 GPa. The calculated bulk moduli show a non-monotonous evolution with the boron content or the a$_0$ lattice parameters. Nonetheless, some bulk modulus measurements were found to be close to those of diamond-like silicon. A type-VIII Na$_8$B$_{4.1}$Si$_{41.9}$ clathrate was isolated, and both single crystal XRD, Raman and FTIR spectroscopy (See Supplementary Material \cite{demoucron_isostructural_2025}) were performed, providing evidence consistent with metallic behavior. The stoichiometry of type-VIII borosilicide clathrates supports the hypothesis that such phases are metallic. At the same time, our attempts to measure electrical resistivity failed due to the inconvenient shape and size of the sample. However, the reflectance measurements over a wide photon energy range can help in such a situation, providing alternative insight into conceptual metallicity, as it was proposed for experimental proof of metallic hydrogen \cite{loubeyre_synchrotron_2020}. These results clarify the pressure-driven atomistic mechanisms in borosilicide clathrates and open future works for tuning their electronic and mechanical properties for functional applications.

\subsection*{Data availability statement}
The data that support the findings of this article are openly available 10.6084/m9.figshare.30647453 \cite{data_iso_demoucron}. The gold pressure calibration, \textit{in situ} XRD for synthesis and the single crystal XRD data can be obtained from the corresponding authors upon 
request.

\begin{acknowledgments} 
We acknowledge assistance from the PSICHE beamline staff of synchrotron SOLEIL (Proposals 20210410–BAG, 20220361-BAG and 20240361-BAG for PSICHE beamline). We acknowledge the European Synchrotron Radiation 
Facility (ESRF) for material support during beamtime at the ID06-LVP beamline under proposal number CH4896 and we acknowledge M. Mezouar for preliminary results acquired on ID27 beamline under proposal number CH4380. We acknowledge the ANR-FRANCE (French National Research Agency) for the financial support of the BCSi project number ANR-21-CE08-0018.
We also thank the support of the high-pressure platform of IMPMC laboratory.
\end{acknowledgments}

\section*{Supplementary Material}

\subsection{Sample synthesis}

\paragraph{Type-I Na$_8$Si$_{46}$ clathrate\\}
The stoichiometric reference Na$_8$Si$_{46}$ type-I (Si-I) clathrate has been prepared at ambient pressure using a classical but recently improved chemical technique, described in the study of \textit{Song et al.} \cite{song_straightforward_2021}. Na$_4$Si$_4$ (4~mmol) was loaded in a pyrolytic h-BN crucible ($\phi$~25~\!$\times$~\!h~60~mm) previously dried at 673~K under vacuum (10$^{-3}$~mbar) for 10~hours. The crucible was inserted in a bottom-closed quartz tube covered with an h-BN cap prior heating, to maintain a high Na vapor pressure. The tube was heated at 743~K under dynamic vacuum (10$^{-3}$~mbar) for 90~min inside a vertical furnace. After dwelling and cooling down, the quartz tube with the h-BN crucible and its containment were transferred into an argon-filled glovebox without exposure to air, and the powder was stored in inert atmosphere.

\paragraph{type-I Na$_8$B$_x$Si$_{46-x}$ borosilicide clathrates\\}
\textit{In-situ} high-pressure high-temperature synthesis at 3.5~GPa and 1150~K of sample composed of type-I borosilicide clathrates was performed using the 20MN Voggenreiter multi-anvil press at beamline ID06-LVP of the ESRF \cite{crichton_ebs_2024}. For this experiment, a mixture of Na$_4$Si$_4$ powder (obtained from the synthesis of reference \cite{song_straightforward_2021}), amorphous boron powder (Sigma-Aldrich, $\geqslant$~\!95~\!\%) and silicon powder (Alfa Aesar, 99.999~\!\%) with the Na:B:Si atomic ratio of 8:8:38 was ground in a ceramic mortar for one hour inside a high-purity Ar glovebox and loaded into a h-BN capsule. We used a 14/8 multianvil assembly (MgO octahedron with 14~mm side compressed with eight WC cubic anvils with 8~mm-side triangular truncations), equipped with graphite furnace and alumina cap to achieve 3.5~GPa and 1150~K. Si-I \cite{bornstein_numerical_1982} and Na \cite{fritz_equation_1984} equations of state were used to calibrate pressure and temperature in parallel with estimates based on previous calibration curves \cite{perdew_self-interaction_1981, perdew_generalized_1996}. Angle dispersion X-ray diffraction was performed using a wavelength corresponding to 33~k$eV$ and collected on an azimuthally-scanning Detection Technology X-Scan c series GOS linear detector. The power was switched off and after that, the pressure was slowly released at the end of the synthesis. 

\paragraph{type-VIII Na$_8$B$_x$Si$_{46-x}$ borosilicide clathrate\\}
A sample made of type-I and type-VIII sodium borosilicide clathrates was obtained from the same mixture described here above. 
The mixture was compressed into a pellet and then introduced into a similar 18/11 multianvil assembly and compressed to 4~GPa. The heater-sample total resistivity probing method was used to achieve the complete transformation of intermediate metallic clathrates \cite{courac_high-pressure_2019}. After the accomplishment of this process, the temperature was gradually decreased over a period of 20~minutes. The latter allowed us to obtain crystallization in (quasi-) equilibrium conditions. The recovered samples were easily removed from the graphite heater. No reaction between the mixture and the graphite capsule or the alumina plugs was detected up to 1500~K at 4~GPa. The type-VIII borosilicide clathrate was found to be the major phase. A small proportion of two type-I clathrate phases were also found inside the same sample powder with a$_0$~=~9.9523(9)~Å and a$_0$~=~10.1190(16)~Å lattice parameters, confirming the presence of boron in their structure. 

\begin{figure}
    \centering
    \includegraphics[width=0.7\linewidth]{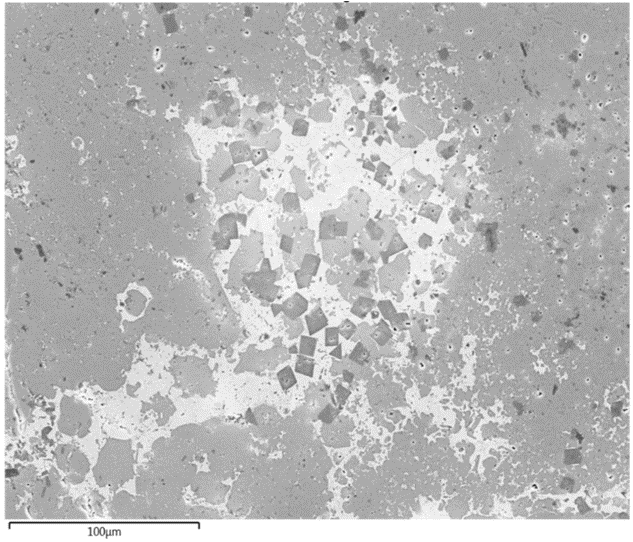}
    \caption{SEM image with AsB detector of the polished sample synthetized at 4~GPa and 1500~K.}
    \label{SM1 SEM}
\end{figure}

Theses mixture of different phase was observed by both XRD and SEM-EDX imaging (See Figure \ref{SM1 SEM}). Small polycrystalline grains, as well as clathrate VIII single crystals of approximately 10~µm could be isolated from the grounded mixture \cite{courac_high-pressure_2019}.

\begin{figure}
    \centering
    \includegraphics[width=0.8\linewidth]{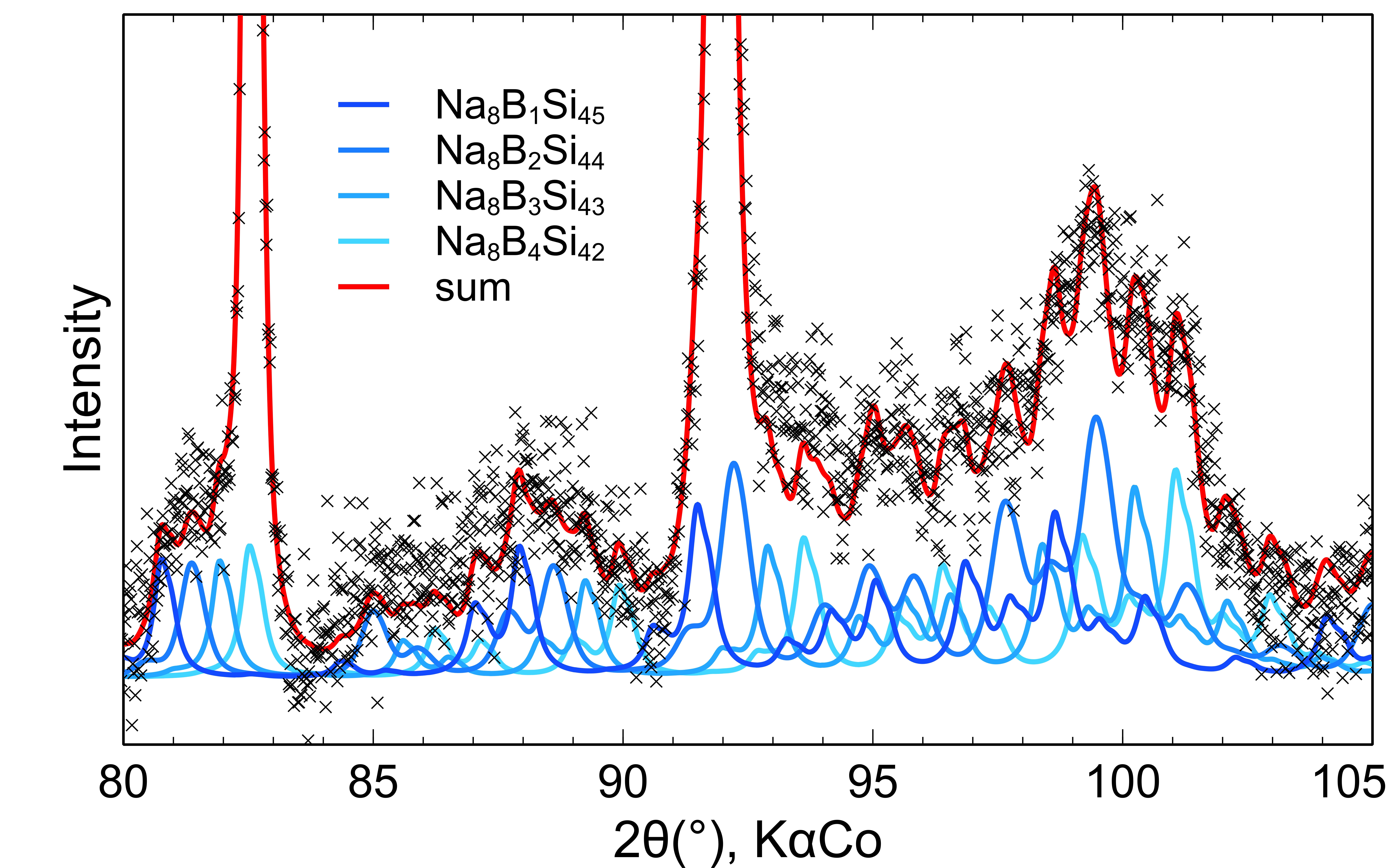}
    \caption{XRD powder pattern  (between 80 and 105°) of the sample synthetized at 3.5~GPa and 1150~K, the intensities of the computed phases are calculated from Na$_8$B$_1$Si$_{45}$ to Na$_8$B$_4$Si$_{42}$ fixed stoichiometry with boron atom inside the 16\textit{i} and 24\textit{k} Wyckoff sites. The intense unlabeled peaks correspond to silicon.}
    \label{SM2 computed 2 NaBSi}
\end{figure}

\begin{figure}
    \centering
    \includegraphics[width=0.8\linewidth]{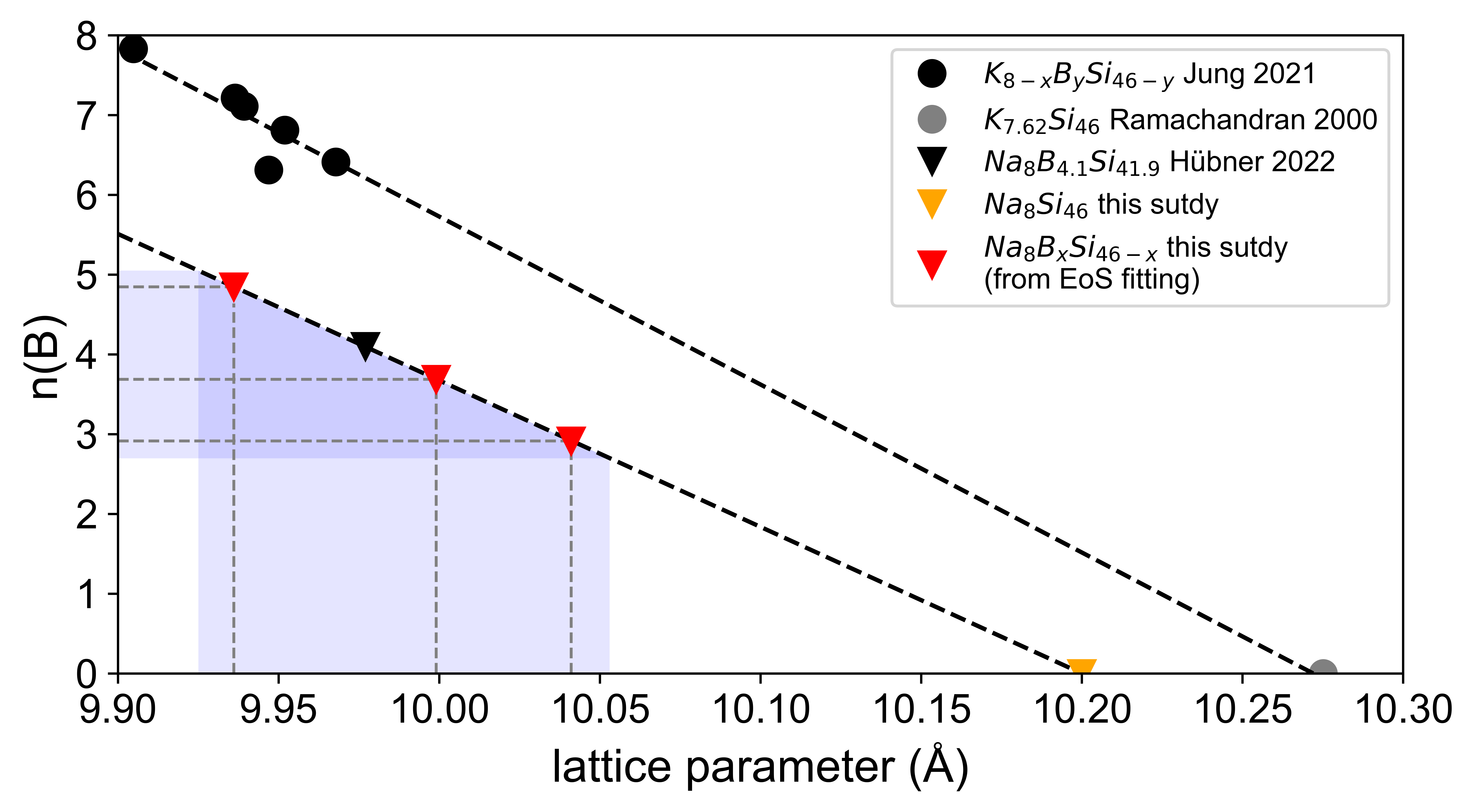}
    \caption{Boron content of K$_8$B$_x$Si$_{46-x}$ and Na$_8$B$_x$Si$_{46-x}$ clathrates in function of the lattice parameter. The  samples synthetized at 5~GPa and 1150~K were placed from the fitted a$_0$ lattice parameters from the EoS.}
    \label{Vegard NaBSi}
\end{figure}

\begin{figure}
    \centering
    \includegraphics[width=1\linewidth]{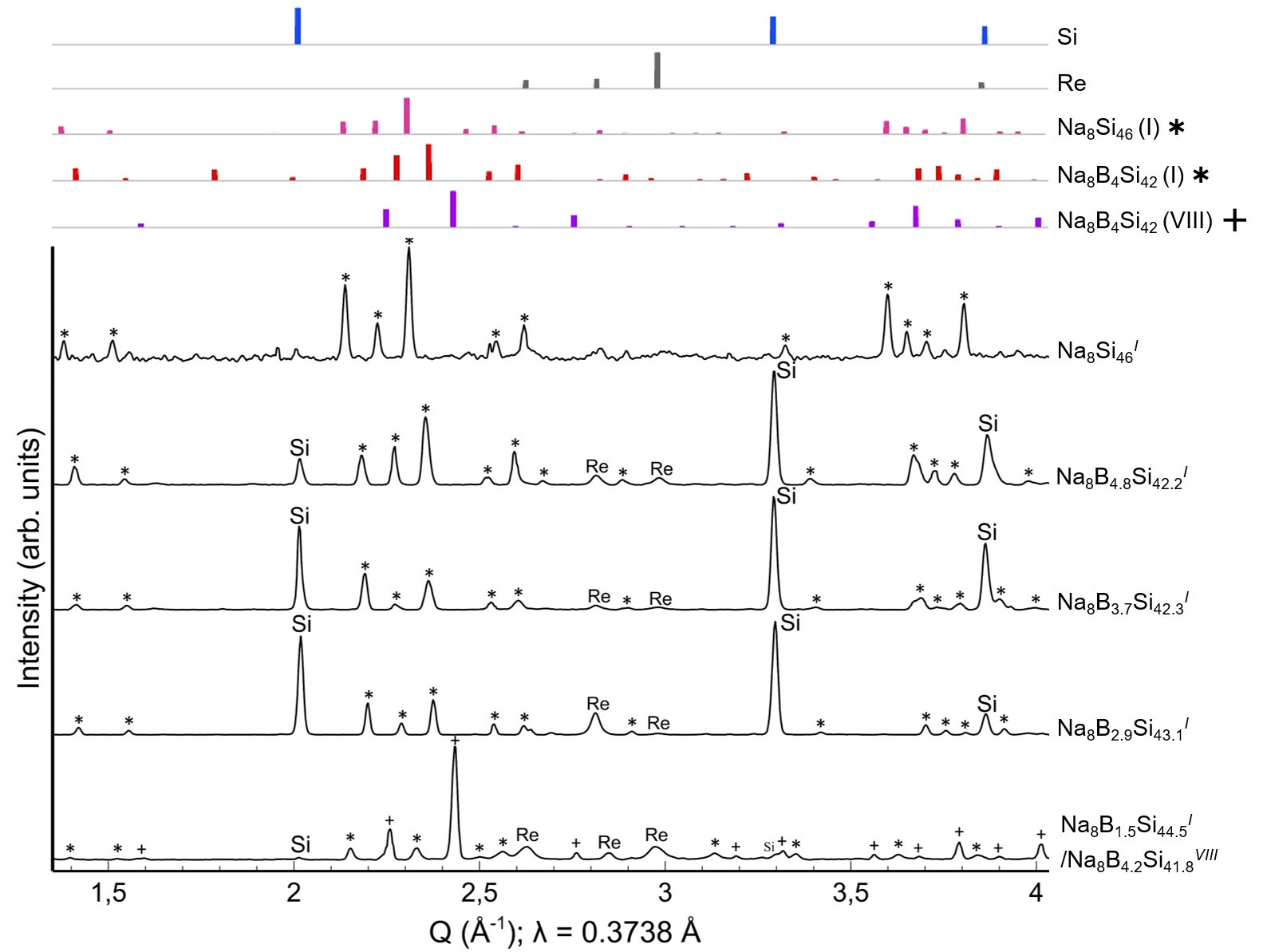}
    \caption{XRD patterns of every Na$_8$B$_x$Si$_{46-x}$ clathrates found at different places in a DAC cell at nearly ambient condition (* : type-I clathrate; + : type-VIII clathrate). The compositions of type-I borosilicides clathrates are establisehd from the Vegard's law presented in Figure \ref{Vegard NaBSi}.  }
    \label{XRD patterns NaBSi}
\end{figure}

\begin{figure}[!t]
    \centering
    \includegraphics[width=1\linewidth]{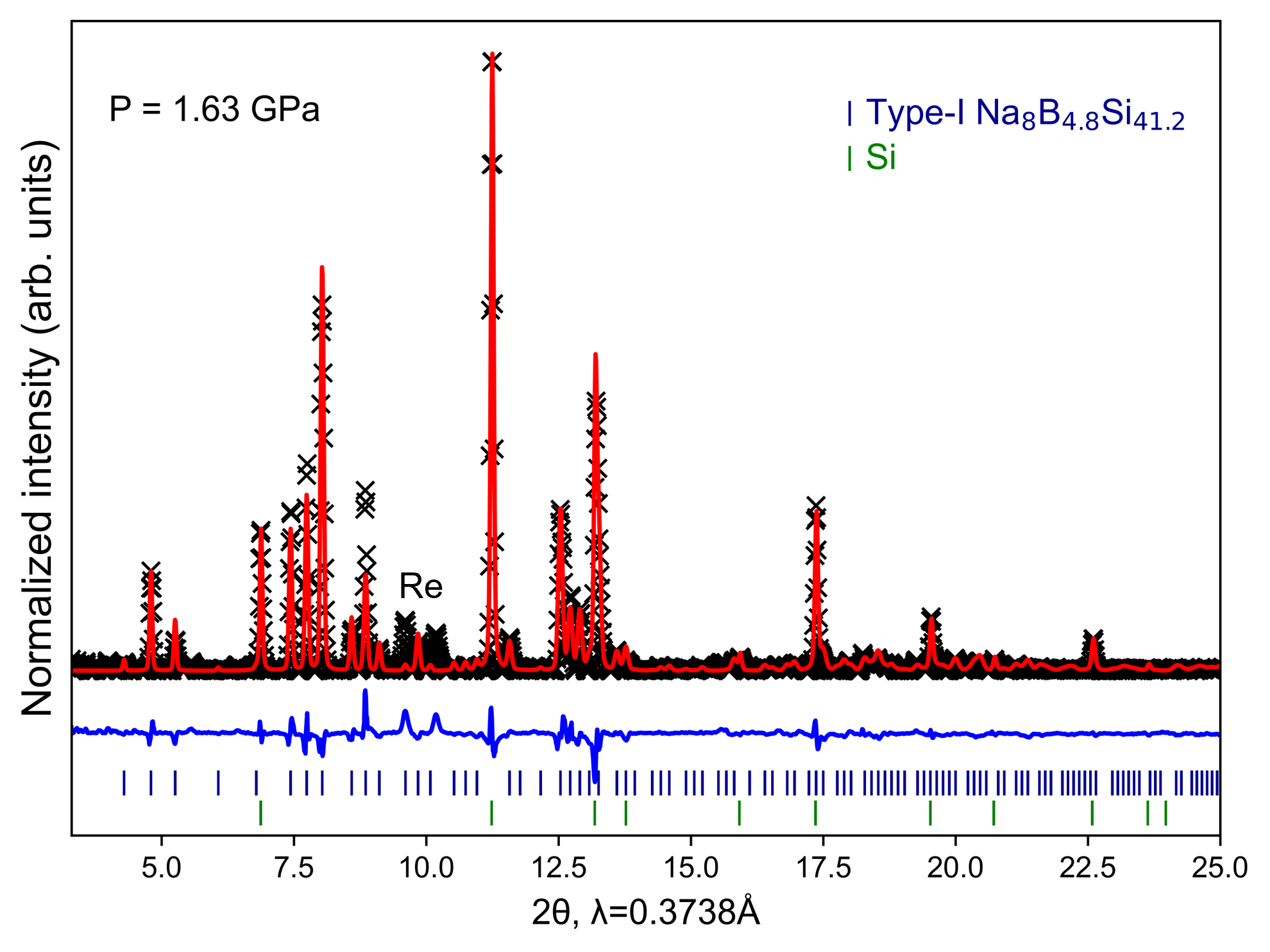}
    \caption{Typical Rietveld refinement obtained during high-pressure \textit{in-situ} experiment, R-factor~=~2.08. Refined parameters : lattice parameters, sites occupancies factors (Si-6\textit{c} and Si-16\textit{i}), microstrain. The silicon refinement does not consider the calculation of the structure factor due to the presence of strong silicon diffraction spots observed on the detector. }
    \label{fig:placeholder}
\end{figure}

\subsection{Pressure calibration}

\begin{figure}
    \centering
    \includegraphics[width=1\linewidth]{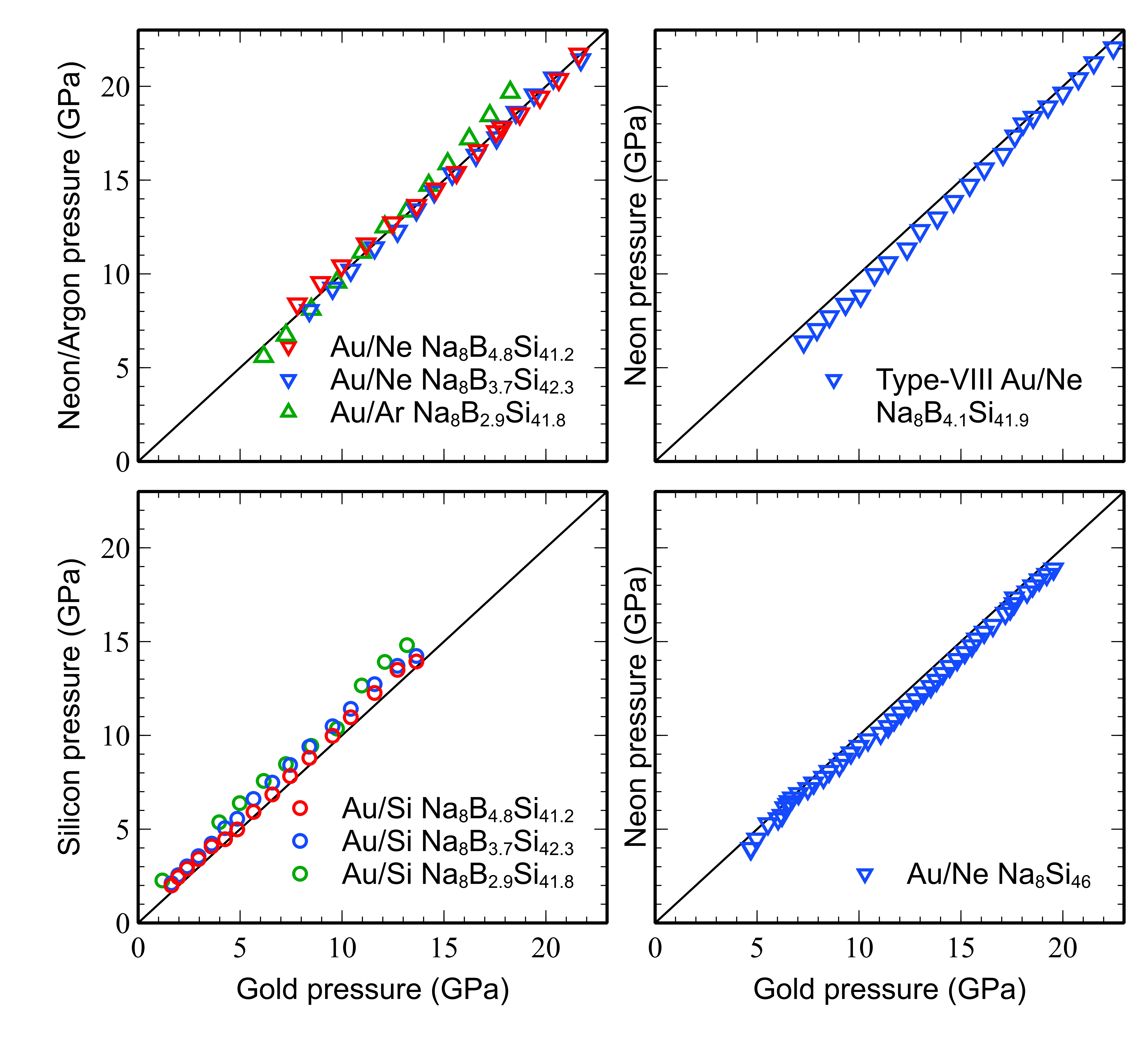}
    \caption{Comparison between the pressures obtained with gold gauge, pressure medium (Ne or Ar, when solidified) and Si grain of sample. No discontinuity at 13~GPa was observed for the different DAC experiments.}
    \label{P vs P NaBSi}
\end{figure}

Since gold and the different samples are not at the same position inside the DAC, the pressure differences between gold and the other pressure calibrants could be explained by the DAC radial pressure gradient \cite{sung_pressure_1977}. Also, the initial volume of the crystallized neon changes according to various equation of states \cite{dewaele_high_2008, anderson_equation_1973, hemley_x-ray_1989} and could explain the pressure differences between gold and neon. However, no significant disruption is observed around 13~GPa, therefore ensuring the reliability of the pressure calibration.

\subsection{Type-VIII borosilicides clathrates characterization}

Single crystal XRD data reported in the following tables come from two different single crystal samples. The two different samples were synthesized at high pressure high temperature (4~GPa and 1500~K) using multi-anvil presses. 
Boron atoms substitute silicon only inside the 8\textit{c} Wyckoff site. Including additional boron substitutional sites in the structural refinement produced no meaningful changes to the fit and these results are, therefore, not reported.
\begin{table*}
\centering
\begin{tabular}{lll}
    ~& Crystal n°1 & Crystal n°2\\ \hline
    Composition& Na$_{8(1)}$B$_{4.1(2)}$Si$_{41.9(2)}$&Na$_8$B$_{4.51(10)}$Si$_{41.49(10)}$\\
    Formula weight (kg.mol$^{-1}$) & 1405.04& 1397.9\\
    Temperature (K) & 293 & 293\\
    Crystal system & Cubic & Cubic\\
    Space group & $I\overline{4}\text{3}m$ & $I\overline{4}\text{3}m$\\
    Lattice parameter (Å) & 9.69910(10) & 9.6794(4) \\
    Volume (Å$^3$) & 912.42(3) & 906.87(6)\\
    Wavelength (Å) & 0.7107& 0.4066\\
    Diffractometer& RIGAKU (Agilent) Xcalibur S& APS, HPCAT\\
    Number of refined reflections & 280& 141\\
    $h,k,l$ index range & -13$\leq$h$\leq$13 ; -13$\leq$k$\leq$13 ; -13$\leq$l$\leq$12 & -12$\leq$h$\leq$11 ; -11$\leq$k$\leq$12 ; -7$\leq$l$\leq$4 \\
    Goodness of fit & 1.241 & 4.81\\
    Final R indexes (I$>$3$\sigma$(I)) & \textit{R(F\textsuperscript{2})} = 0.0438, \textit{wR(F\textsuperscript{2})} = 0.1073 & \textit{R(F\textsuperscript{2})} = 0.0599, \textit{wR(F\textsuperscript{2})} = 0.0740\\
    Software for refinement & Olex2, Shelx & Olex2, Shelx \\  \hline
\end{tabular}
\caption{Type-VIII Na$_8$B$_{4.1(2)}$Si$_{41.9(2)}$ and type VIII Na$_8$B$_{4.51(10)}$Si$_{41.49(10)}$ experimental crystallographic data processing.}
\label{SM tab 1 type VIII}
\end{table*}

\begin{table*}
\centering
\begin{tabular}{ccccccc}
\hline
\textbf{Atom} & \textbf{Site} & \textbf{x} & \textbf{y} & \textbf{z} & \textbf{B (temp)} & \textbf{Occupancy} \\ \hline
Na1 & 8c & 0.6882(2) & 0.6882(2) & 0.6882(2) & 1.0106 & 1 \\ \hline
Si1 & 12\textit{d} & 0.25 & 0.5 & 0 & 0.1921 & 1 \\ 
Si2 & 2\textit{a} & 0 & 0 & 0 & 0.4974 & 1 \\ 
Si3 & 24\textit{g} & 0.08527(17) & 0.08527(13) & 0.65559(13) & 0.3606 & 1 \\ 
\multirow{2}{*}{Si4/B} & \multirow{2}{*}{8\textit{c}} & \multirow{2}{*}{0.8654(2)} & \multirow{2}{*}{0.8654(2)} & \multirow{2}{*}{0.8654(2)} & \multirow{2}{*}{0.5922} & 0.49(2) (Si) \\ 
  &   &   &   &   &   & 0.51(2) (B) \\ \hline
\end{tabular}
\caption{Experimental structure data of Na$_8$B$_{4.1(2)}$Si$_{41.9(2)}$ type-VIII clathrate. The associated anisotropic displacement tensors U$_{ij}$ are: \\
Na1 : U$_{11}$~\!=~\!U$_{22}$~\!=~\!U$_{33}$~\!=~\!0.0128(9) ; U$_{23}$~\!=~\!U$_{13}$~\!=~\!U$_{12}$~\!=~\!0.0046(9) \\
Si1 : U$_{11}$~\!=~\!U$_{22}$~\!=~\!U$_{33}$~\!=~\!0.0075(13) ; U$_{23}$~\!=~\!U$_{13}$~\!=~\!U$_{12}$~\!=~\!0.0030(8) \\
Si2 : U$_{11}$~\!=~\!U$_{22}$~\!=~\!U$_{33}$~\!=~\!0.0063(9) ; U$_{23}$~\!=~\!U$_{13}$~\!=~\!U$_{12}$~\!=~\!0 \\
Si3 : U$_{11}$~\!=~\!0.0101(8) ; U$_{22}$~\!=~\!U$_{33}$~\!=~\!0.0018(5) ; U$_{23}$~\!=~\!0.0013(5) ; U$_{13}$~\!=~\!U$_{12}$~\!=~\!0.0018(3) \\
Si4/B : U$_{11}$ ~\!=~\!U$_{33}$~\!=~\!0.0161(4) ; U$_{22}$~\!=~\!0.0037(8) ; U$_{23}$~\!=~\!U$_{13}$~\!=~\!U$_{12}$~\!=~\!0}
\label{SM tab 2 type VIII}
\end{table*}

\begin{table*}
\centering
\begin{tabular}{ccccccc}
\hline
\textbf{Atom} & \textbf{Site} & \textbf{x} & \textbf{y} & \textbf{z} & \textbf{B (temp)} & \textbf{Occupancy} \\ \hline
Na1 & 8\textit{c} & 0.18798(14) & 0.18798(14) & 0.18798(14) & 0.0229(5) & 1 \\ 
Si1 & 12\textit{d} & 0.25 & 0.5 & 0 & 0.0161(4) & 1 \\ 
Si2 & 2\textit{a} & 0 & 0 & 0 & 0.0115(2) & 1 \\ 
Si3 & 24\textit{g} & 0.41438(6) & 0.41438(6) & 0.15688(8) & 0.0133(19) & 1 \\ 
\multirow{2}{*}{Si4/B} & \multirow{2}{*}{8\textit{c}} & \multirow{2}{*}{0.52294(13)} & \multirow{2}{*}{0.52294(13)} & \multirow{2}{*}{0.52294(13)} & \multirow{2}{*}{0.0154(6)} & 0.436(12) (Si) \\ 
  &   &   &   &   &   & 0.564(12) (B) \\ \hline
\end{tabular}
\caption{Experimental structure data of Na$_8$B$_{4.51(10)}$Si$_{41.49(10)}$ type-VIII clathrate. The associated anisotropic displacement tensors U$_{ij}$ are: \\
Na1 : U$_{11}$~\!=~\!U$_{22}$~\!=~\!U$_{33}$~\!=~\!0.0229(5) ; U$_{23}$~\!=~\!U$_{13}$~\!=~\!U$_{12}$~\!=~\!-0.0037(5) \\
Si1 : U$_{11}$~\!=~\!U$_{22}$~\!=~\!U$_{33}$~\!=~\!0.0161(4) ; U$_{23}$~\!=~\!U$_{13}$~\!=~\!U$_{12}$~\!=~\!0\\
Si2 : U$_{11}$~\!=~\!0.0120(3) ; U$_{22}$~\!=~\!U$_{33}$~\!=~\!0.0112(3) ; U$_{23}$~\!=~\!U$_{13}$~\!=~\!U$_{12}$~\!=~\!0 \\
Si3 : U$_{11}$~\!=~\!U$_{22}$~\!=~\!0.0106(2) ; U$_{33}$~\!=~\!0.0187(3) ; U$_{23}$~\!=~\!U$_{13}$~\!=~\!0.00213(16) ; U$_{12}$~\!=~\!0.0012(2) \\
Si4/B : U$_{11}$~\!=~\!U$_{22}$~\!=~\!U$_{33}$~\!=~\!0.0154(6) ; U$_{23}$~\!=~\!U$_{13}$~\!=~\!U$_{12}$~\!=~\!-0.0026(4)}
\label{SM tab 4 type VIII}
\end{table*}


\begin{figure}
    \centering
    \includegraphics[width=1\linewidth]{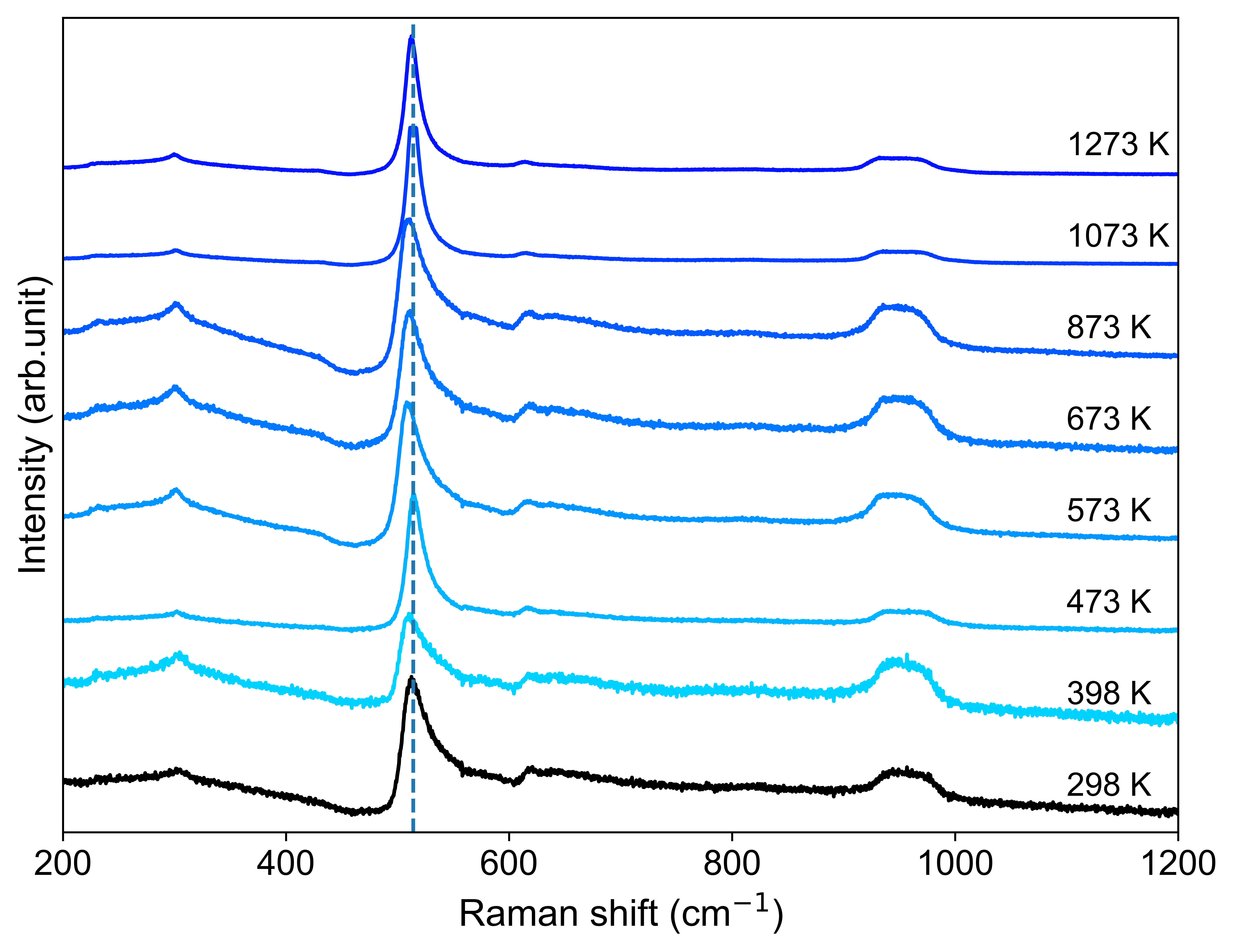}
    \caption{Raman spectrum of Na$_8$B$_{4.51}$Si$_{41.49}$ type-VIII single-crystal clathrate after annealing at high temperature.}
    \label{Raman HT NaBSi}
\end{figure}

\begin{figure}
    \centering
    \includegraphics[width=1\linewidth]{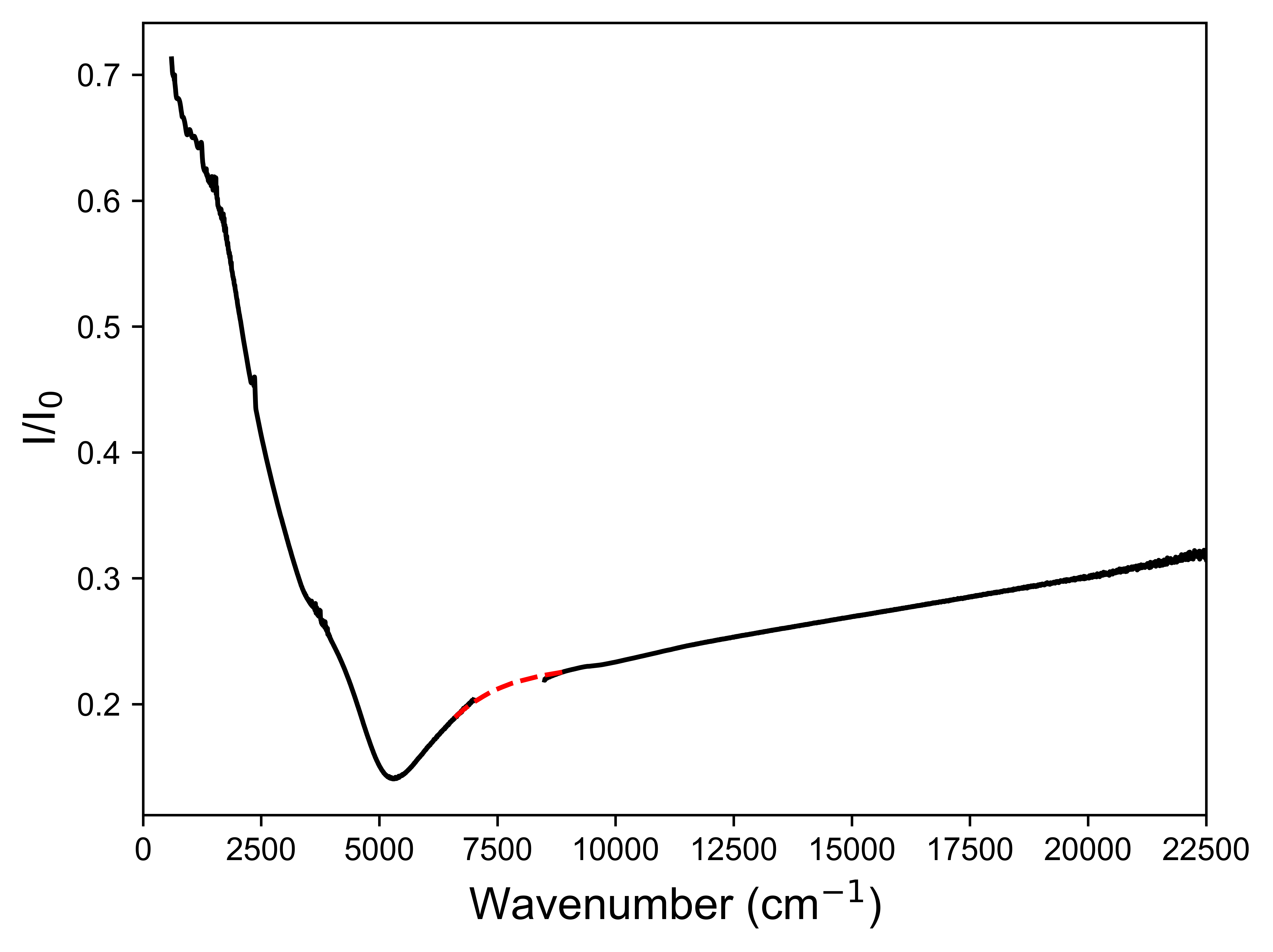}
    \caption{Reflectance FTIR spectroscopy at ambient condition of Na$_8$B$_{4.51}$Si$_{41.49}$ type-VIII single-crystal clathrate.}
    \label{FTIR}
\end{figure}

The FTIR reflectance spectrum exhibits a trend similar to that of typical metals (Al, Ag, Au) \cite{shanks_optics_2016} as well as to that of the BC8-Si (Si-III) phase, which is classified as a narrow-gap semiconductor \cite{zhang_bc8_2017}. The Na$_8$B$_{4.51}$Si$_{41.49}$ type-VIII single-crystal clathrate shows a reflectance minimum at 5295~cm$^{-1}$, where the I/I$_0$ ratio drops to 0.14. This minimum corresponds to an absorption energy of 0.66~eV, which lies between the absorption energies of BC8-Si (0.16~eV) and of the other metals ($>$~2~eV). Additional resistivity measurements are therefore required to confirm the metallicity of this phase, while the hypothesis of a narrow-gap semiconductor is not entirely excluded.

\clearpage
\section*{References}
\bibliography{apssamp}

@article{kishimoto_synthesis_2008,
	title = {Synthesis and Thermoelectric Properties of Silicon Clathrates {Sr}$_8${Al}$_x${Ga}$_{16-x}${Si}$_{30}$ with the Type-{I} and Type-{VIII} Structures},
	volume = {1},
	issn = {1882-0786},
	url = {https://iopscience.iop.org/article/10.1143/APEX.1.031201},
	doi = {10.1143/APEX.1.031201},
	pages = {031201},
	number = {3},
	journal = {Applied Physics Express},
	shortjournal = {Appl. Phys. Express},
	author = {Kishimoto, Kengo and Ikeda, Naoya and Akai, Koji and Koyanagi, Tsuyoshi},
	urldate = {2025-12-17},
	date = {2008-03-14},
    year = {2008},
	langid = {english},
	note = {Publisher: {IOP} Publishing},
}

@article{suekuni_simultaneous_2008,
	title = {Simultaneous structure and carrier tuning of dimorphic clathrate Ba 8 Ga 16 Sn 30},
	volume = {77},
	rights = {http://link.aps.org/licenses/aps-default-license},
	issn = {1098-0121, 1550-235X},
	url = {https://link.aps.org/doi/10.1103/PhysRevB.77.235119},
	doi = {10.1103/PhysRevB.77.235119},
	pages = {235119},
	number = {23},
	journal = {Physical Review B},
	shortjournal = {Phys. Rev. B},
	author = {Suekuni, K. and Avila, M. A. and Umeo, K. and Fukuoka, H. and Yamanaka, S. and Nakagawa, T. and Takabatake, T.},
	urldate = {2025-12-16},
	date = {2008-06-25},
    year = {2008},
	langid = {english},
}

@article{sasaki_synthesis_2009,
	title = {Synthesis and thermoelectric properties of type-{VIII} germanium clathrates {Sr}$_8${Al}$_x${Ga}$_y${Ge}$_{46-x-y}$},
	volume = {105},
	issn = {0021-8979, 1089-7550},
	url = {https://pubs.aip.org/jap/article/105/7/073702/396759/Synthesis-and-thermoelectric-properties-of-type},
	doi = {10.1063/1.3100205},
	pages = {073702},
	number = {7},
	journal = {Journal of Applied Physics},
	author = {Sasaki, Yuta and Kishimoto, Kengo and Koyanagi, Tsuyoshi and Asada, Hironori and Akai, Koji},
	urldate = {2025-12-16},
	date = {2009-04-01},
    year = {2009},
	langid = {english},
}

@article{toulemonde_high_2011,
	title = {High pressure x-ray diffraction study of the volume collapse in {Ba}$_{24}${Si}$_{100}$ clathrate},
	volume = {83},
	rights = {http://link.aps.org/licenses/aps-default-license},
	issn = {1098-0121, 1550-235X},
	url = {https://link.aps.org/doi/10.1103/PhysRevB.83.134110},
	doi = {10.1103/PhysRevB.83.134110},
	pages = {134110},
	number = {13},
	journal = {Physical Review B},
	shortjournal = {Phys. Rev. B},
	author = {Toulemonde, P. and Machon, D. and San Miguel, A. and Amboage, M.},
	urldate = {2025-12-01},
	date = {2011-04-08},
    year = {2011},
	langid = {english},
}

@misc{data_iso_demoucron,
	title = {Data Isostructural phase transition and equation of state of type-{I} and -{VIII} metallic sodium borosilicide clathrates, https://figshare.com/account/articles/30647453},
	url = {https://figshare.com/account/articles/30647453},
    author = {Demoucron, M},
	urldate = {2025-11-27},
	publisher = {figshare},
    year = {2025}
}

@article{demoucron_isostructural_2025,
    author = {Demoucron, M and Pandolfi, S and Guarnelli, Y and Baptiste, B and Chevignon, P and Guignot, N and  Portehault, D and Strobel, T A and Crichton, W A and Le Godec, Y and Courac, A},
    title = {Supplementary material, Isostructural phase transition and equation of state of type-I and type-VIII metallic sodium borosilicide clathrates} ,
    journal = {Physical Review Letters},
    year = {2025},
}

@incollection{shevelkov_zintl_2011,
	address = {Berlin, Heidelberg},
	title = {Zintl Clathrates},
	isbn = {978-3-642-21150-8},
	url = {https://doi.org/10.1007/430_2010_25},
	urldate = {2025-11-24},
	booktitle = {Zintl Phases: Principles and Recent Developments},
	publisher = {Springer},
	author = {Shevelkov, Andrei V. and Kovnir, Kirill},
	editor = {Fässler, Thomas F.},
	year = {2011},
	doi = {10.1007/430_2010_25},
	pages = {97--142},
}

@article{zhang_bc8_2017,
	title = {{BC}8 Silicon (Si-{III}) is a Narrow-Gap Semiconductor},
	volume = {118},
	url = {https://link.aps.org/doi/10.1103/PhysRevLett.118.146601},
	doi = {10.1103/PhysRevLett.118.146601},
	number = {14},
	urldate = {2025-11-19},
	journal = {Physical Review Letters},
	author = {Zhang, Haidong and Liu, Hanyu and Wei, Kaya and Kurakevych, Oleksandr O. and Le Godec, Yann and Liu, Zhenxian and Martin, Joshua and Guerrette, Michael and Nolas, George S. and Strobel, Timothy A.},
	month = apr,
	year = {2017},
	note = {Publisher: American Physical Society},
	pages = {146601},
}

@article{yamanaka_high-pressure_2014,
	title = {High-Pressure Synthesis and Structural Characterization of the Type {II} Clathrate Compound {Na}$_{\textrm{30.5}}$ {Si}$_{\textrm{136}}$ Encapsulating Two Sodium Atoms in the Same Silicon Polyhedral Cages},
	volume = {136},
	issn = {0002-7863, 1520-5126},
	url = {https://pubs.acs.org/doi/10.1021/ja502733e},
	doi = {10.1021/ja502733e},
	pages = {7717--7725},
	number = {21},
	journal = {Journal of the American Chemical Society},
	shortjournal = {J. Am. Chem. Soc.},
	author = {Yamanaka, Shoji and Komatsu, Masaya and Tanaka, Masashi and Sawa, Hiroshi and Inumaru, Kei},
	urldate = {2025-11-17},
	date = {2014-05-28},
    year = {2014},
	langid = {english},
}

@article{kurakevych_na-si_2013,
	title = {Na-Si Clathrates Are High-Pressure Phases: A Melt-Based Route to Control Stoichiometry and Properties},
	volume = {13},
	issn = {1528-7483, 1528-7505},
	url = {https://pubs.acs.org/doi/10.1021/cg3017084},
	doi = {10.1021/cg3017084},
	shorttitle = {Na-Si Clathrates Are High-Pressure Phases},
	pages = {303--307},
	number = {1},
	journal = {Crystal Growth \& Design},
	shortjournal = {Crystal Growth \& Design},
	author = {Kurakevych, Oleksandr O. and Strobel, Timothy A. and Kim, Duck Young and Muramatsu, Takaki and Struzhkin, Viktor V.},
	urldate = {2023-10-16},
	date = {2013-01-02},
    year = {2013},
	langid = {english},
}

@article{shanks_optics_2016,
	title = {Optics for concentrating photovoltaics: Trends, limits and opportunities for materials and design},
	volume = {60},
	issn = {13640321},
	url = {https://linkinghub.elsevier.com/retrieve/pii/S1364032116001192},
	doi = {10.1016/j.rser.2016.01.089},
	shorttitle = {Optics for concentrating photovoltaics},
	pages = {394--407},
	journal = {Renewable and Sustainable Energy Reviews},
	shortjournal = {Renewable and Sustainable Energy Reviews},
	author = {Shanks, Katie and Senthilarasu, S. and Mallick, Tapas K.},
	urldate = {2025-11-17},
	date = {2016-07},
    year = {2016},
	langid = {english},
}

@article{rani_fano-type_2024,
	title = {Fano-type discrete-continuum interaction in perovskites and its manifestation in Raman spectral line shapes},
	volume = {60},
	issn = {1359-7345, 1364-548X},
	url = {https://xlink.rsc.org/?DOI=D3CC05789E},
	doi = {10.1039/D3CC05789E},
	pages = {2115--2124},
	number = {16},
	journal = {Chemical Communications},
	shortjournal = {Chem. Commun.},
	author = {Rani, Chanchal and Kumar, Rajesh},
	urldate = {2025-11-17},
	date = {2024},
    year = {2024},
	langid = {english},
}

@article{nesper_zintl-klemm_2014,
	title = {The Zintl-Klemm Concept – A Historical Survey},
	volume = {640},
	rights = {Copyright © 2014 {WILEY}-{VCH} Verlag {GmbH} \& Co. {KGaA}, Weinheim},
	issn = {1521-3749},
	url = {https://onlinelibrary.wiley.com/doi/abs/10.1002/zaac.201400403},
	doi = {10.1002/zaac.201400403},
	pages = {2639--2648},
	number = {14},
	journal = {Zeitschrift für anorganische und allgemeine Chemie},
	author = {Nesper, Reinhard},
	urldate = {2025-09-03},
	date = {2014},
    year = {2014},
	langid = {english},
}

@article{kumar_effect_2021,
	title = {Effect of some physical perturbations and their interplay on Raman spectral line shapes in silicon: A brief review},
	volume = {52},
	rights = {© 2021 John Wiley \& Sons, Ltd.},
	issn = {1097-4555},
	url = {https://onlinelibrary.wiley.com/doi/abs/10.1002/jrs.6272},
	doi = {10.1002/jrs.6272},
	shorttitle = {Effect of some physical perturbations and their interplay on Raman spectral line shapes in silicon},
	pages = {2100--2118},
	number = {12},
	journal = {Journal of Raman Spectroscopy},
	author = {Kumar, Rajesh and Tanwar, Manushree},
	urldate = {2025-07-21},
	date = {2021},
    year = {2021},
	langid = {english},
}

@article{kasper_clathrate_1965,
	title = {Clathrate Structure of Silicon {Na}$_8$Si$_{46}$ and {Na}$_x${Si}$_{136}$ (x{\textless}11)},
	volume = {150},
	url = {https://www.science.org/doi/10.1126/science.150.3704.1713},
	doi = {10.1126/science.150.3704.1713},
	pages = {1713--1714},
	number = {3704},
	journal = {Science},
	author = {Kasper, John S. and Hagenmuller, Paul and Pouchard, Michel and Cros, Christian},
	urldate = {2025-07-22},
	date = {1965-12-24},
    year = {1965},
	note = {Publisher: American Association for the Advancement of Science},
}

@misc{lutterotti_maud_2025,
	title = {MAUD : Materials Analysis Using Diffraction},
	version = {2.9998},
	author = {Lutterotti, L.},
	date = {2025},
    year = {2025},
}

@article{vinet_compressibility_1987,
	title = {Compressibility of solids},
	volume = {92},
	issn = {2156-2202},
	url = {https://onlinelibrary.wiley.com/doi/abs/10.1029/JB092iB09p09319},
	doi = {10.1029/JB092iB09p09319},
	pages = {9319--9325},
	issue = {B9},
	journal = {Journal of Geophysical Research: Solid Earth},
	author = {Vinet, P. and Ferrante, J. and Rose, J. H. and Smith, J. R.},
	urldate = {2022-12-12},
	date = {1987},
    year = {1987},
	langid = {english},
}

@article{hubner_mastering_2021,
	title = {Mastering extreme size constraints in the clathrate-I borosilicide {Cs}$_8${B}$_8${Si}$_{38}$},
	volume = {647},
	issn = {1521-3749},
	url = {https://onlinelibrary.wiley.com/doi/abs/10.1002/zaac.202000400},
	doi = {10.1002/zaac.202000400},
	pages = {119--125},
	number = {2},
	journal = {Zeitschrift für anorganische und allgemeine Chemie},
	author = {Hübner, Julia-Maria and Jung, Walter and Koželj, Primož and Bobnar, Matej and Cardoso-Gil, Raul and Burkhardt, Ulrich and Böhme, Bodo and Baitinger, Michael and Schwarz, Ulrich and Grin, Yuri},
	urldate = {2024-12-12},
	date = {2021},
    year = {2021},
	langid = {english},
}

@article{crichton_ebs_2024,
	title = {The {EBS} large-volume press facility at the {ESRF}},
	volume = {80},
	issn = {2053-2733},
	url = {https://journals.iucr.org/paper?S2053273324095846},
	doi = {10.1107/S2053273324095846},
	pages = {e415--e415},
	issue = {a1},
	journal = {Acta Crystallographica Section A Foundations and Advances},
	shortjournal = {Acta Crystallogr A Found Adv},
	author = {Crichton, Wilson A.},
	urldate = {2025-05-09},
	date = {2024-08-26},
    year = {2024},
	langid = {english},
}

@article{bornstein_numerical_1982,
	title = {Numerical data and functional relationships in science and technology, group iii: crystal and solid state physics, Vol. 13: Metals: phonon states, electron states and fermi surfaces, subvolume a, phonon states of elements, electron states and fermi surfaces of alloys},
	volume = {17},
	rights = {Copyright © 1982 {WILEY}-{VCH} Verlag {GmbH} \& Co. {KGaA}},
	issn = {1521-4079},
	url = {https://onlinelibrary.wiley.com/doi/abs/10.1002/crat.2170170310},
	doi = {10.1002/crat.2170170310},
	shorttitle = {Numerical data and functional relationships in science and technology, group iii},
	pages = {326--326},
	number = {3},
	journal = {Crystal Research and Technology},
	author = {Börnstein, Landolt},
	urldate = {2025-05-09},
	date = {1982},
    year = {1982},
	langid = {english},
}

@article{fritz_equation_1984,
	title = {Equation of state of sodium},
	volume = {80},
	issn = {0021-9606, 1089-7690},
	url = {https://pubs.aip.org/jcp/article/80/6/2864/154241/Equation-of-state-of-sodiumEquation-of-state-of},
	doi = {10.1063/1.447035},
	pages = {2864--2871},
	number = {6},
	journal = {The Journal of Chemical Physics},
	author = {Fritz, J. N. and Olinger, Bart},
	urldate = {2025-05-09},
	date = {1984-03-15},
    year = {1984},
	langid = {english},
}

@article{perdew_generalized_1996,
	title = {Generalized Gradient Approximation Made Simple},
	volume = {77},
	url = {https://link.aps.org/doi/10.1103/PhysRevLett.77.3865},
	doi = {10.1103/PhysRevLett.77.3865},
	pages = {3865--3868},
	number = {18},
	journal = {Physical Review Letters},
	shortjournal = {Phys. Rev. Lett.},
	author = {Perdew, John P. and Burke, Kieron and Ernzerhof, Matthias},
	urldate = {2025-05-09},
	date = {1996-10-28},
    year = {1996},
	note = {Publisher: American Physical Society},
}

@article{perdew_self-interaction_1981,
	title = {Self-interaction correction to density-functional approximations for many-electron systems},
	volume = {23},
	rights = {http://link.aps.org/licenses/aps-default-license},
	issn = {0163-1829},
	url = {https://link.aps.org/doi/10.1103/PhysRevB.23.5048},
	doi = {10.1103/PhysRevB.23.5048},
	pages = {5048--5079},
	number = {10},
	journal = {Physical Review B},
	shortjournal = {Phys. Rev. B},
	author = {Perdew, J. P. and Zunger, Alex},
	urldate = {2025-05-09},
	date = {1981-05-15},
    year = {1981},
	langid = {english},
}

@article{anzellini_quasi-hydrostatic_2019,
	title = {Quasi-hydrostatic equation of state of silicon up to 1 megabar at ambient temperature},
	volume = {9},
	issn = {2045-2322},
	url = {https://www.ncbi.nlm.nih.gov/pmc/articles/PMC6820762/},
	doi = {10.1038/s41598-019-51931-1},
	pages = {15537},
	journal = {Scientific Reports},
	shortjournal = {Sci Rep},
	author = {Anzellini, Simone and Wharmby, Michael T. and Miozzi, Francesca and Kleppe, Annette and Daisenberger, Dominik and Wilhelm, Heribert},
	urldate = {2022-12-08},
	date = {2019-10-29},
    year = {2019},
	pmid = {31664104},
	pmcid = {PMC6820762},
}

@article{hemley_x-ray_1989,
	title = {X-ray diffraction and equation of state of solid neon to 110 {GPa}},
	volume = {39},
	issn = {0163-1829},
	url = {https://link.aps.org/doi/10.1103/PhysRevB.39.11820},
	doi = {10.1103/PhysRevB.39.11820},
	pages = {11820--11827},
	number = {16},
	journal = {Physical Review B},
	shortjournal = {Phys. Rev. B},
	author = {Hemley, R. J. and Zha, C. S. and Jephcoat, A. P. and Mao, H. K. and Finger, L. W. and Cox, D. E.},
	urldate = {2022-12-15},
	date = {1989-06-01},
    year = {1989},
	langid = {english},
}

@article{hubner_borosilicide_2022,
	title = {A Borosilicide with Clathrate {VIII} Structure},
	volume = {144},
	issn = {0002-7863, 1520-5126},
	url = {https://pubs.acs.org/doi/10.1021/jacs.2c04745},
	doi = {10.1021/jacs.2c04745},
	pages = {13456--13460},
	number = {30},
	journal = {Journal of the American Chemical Society},
	shortjournal = {J. Am. Chem. Soc.},
	author = {Hübner, Julia-Maria and Carrillo-Cabrera, Wilder and Kozelj, Primoz and Prots, Yurii and Baitinger, Michael and Schwarz, Ulrich and Jung, Walter},
	urldate = {2023-02-06},
	date = {2022-08-03},
    year = {2022},
	langid = {english},
}

@article{dopilka_structural_2021,
	title = {Structural Origin of Reversible Li Insertion in Guest-Free, Type-{II} Silicon Clathrates},
	volume = {2},
	issn = {2699-9412},
	url = {https://onlinelibrary.wiley.com/doi/abs/10.1002/aesr.202000114},
	doi = {10.1002/aesr.202000114},
	pages = {2000114},
	number = {5},
	journal = {Advanced Energy and Sustainability Research},
	author = {Dopilka, Andrew and Weller, J. Mark and Ovchinnikov, Alexander and Childs, Amanda and Bobev, Svilen and Peng, Xihong and Chan, Candace K.},
	urldate = {2023-03-01},
	date = {2021},
    year = {2021},
	langid = {english},
}

@article{dolyniuk_clathrate_2016,
	title = {Clathrate thermoelectrics},
	volume = {108},
	issn = {0927796X},
	url = {https://linkinghub.elsevier.com/retrieve/pii/S0927796X16300237},
	doi = {10.1016/j.mser.2016.08.001},
	pages = {1--46},
	journal = {Materials Science and Engineering: R: Reports},
	shortjournal = {Materials Science and Engineering: R: Reports},
	author = {Dolyniuk, Juli-Anna and Owens-Baird, Bryan and Wang, Jian and Zaikina, Julia V. and Kovnir, Kirill},
	urldate = {2023-03-01},
	date = {2016-10},
    year = {2016},
	langid = {english},
}

@article{dewaele_stability_2021,
	title = {Stability and equation of state of face-centered cubic and hexagonal close packed phases of argon under pressure},
	volume = {11},
	issn = {2045-2322},
	url = {https://www.nature.com/articles/s41598-021-93995-y},
	doi = {10.1038/s41598-021-93995-y},
	pages = {15192},
	number = {1},
	journal = {Scientific Reports},
	shortjournal = {Sci Rep},
	author = {Dewaele, Agnès and Rosa, Angelika D. and Guignot, Nicolas and Andrault, Denis and Rodrigues, João Elias F. S. and Garbarino, Gaston},
	urldate = {2023-10-31},
	date = {2021-07-26},
    year = {2021},
	langid = {english},
}

@article{anderson_anharmonicity_1989,
	title = {Anharmonicity and the equation of state for gold},
	volume = {65},
	issn = {0021-8979, 1089-7550},
	url = {https://pubs.aip.org/jap/article/65/4/1534/176003/Anharmonicity-and-the-equation-of-state-for-gold},
	doi = {10.1063/1.342969},
	pages = {1534--1543},
	number = {4},
	journal = {Journal of Applied Physics},
	author = {Anderson, Orson L. and Isaak, Donald G. and Yamamoto, Shigeru},
	urldate = {2023-11-15},
	date = {1989-02-15},
    year = {1989},
	langid = {english},
}

@article{dewaele_high_2008,
	title = {High pressure–high temperature equations of state of neon and diamond},
	volume = {77},
	issn = {1098-0121, 1550-235X},
	url = {https://link.aps.org/doi/10.1103/PhysRevB.77.094106},
	doi = {10.1103/PhysRevB.77.094106},
	pages = {094106},
	number = {9},
	journal = {Physical Review B},
	shortjournal = {Phys. Rev. B},
	author = {Dewaele, Agnès and Datchi, Frédéric and Loubeyre, Paul and Mezouar, Mohamed},
	urldate = {2023-11-16},
	date = {2008-03-06},
    year = {2008},
	langid = {english},
}

@article{mcmahon_pressure_1994,
	title = {Pressure dependence of the \textit{Imma} phase of silicon},
	volume = {50},
	issn = {0163-1829, 1095-3795},
	url = {https://link.aps.org/doi/10.1103/PhysRevB.50.739},
	doi = {10.1103/PhysRevB.50.739},
	pages = {739--743},
	number = {2},
	journal = {Physical Review B},
	shortjournal = {Phys. Rev. B},
	author = {{McMahon}, M. I. and Nelmes, R. J. and Wright, N. G. and Allan, D. R.},
	urldate = {2023-11-24},
	date = {1994-07-01},
    year = {1994},
	langid = {english},
}

@article{leoni_modelling_2003,
	title = {Modelling of the $\alpha$ (clathrate {VIII}) $\Leftrightarrow$ $\beta$ (clathrate {I}) phase transition in {Eu}$_8${Ga}$_{16}${Ge}$_{30}$},
	journal = {Journal of Alloys and Compounds},
	author = {Leoni, Stefano and Carrillo-Cabrera, Wilder and Grin, Yuri},
	date = {2003},
    year = {2003},
	langid = {english},
}

@article{neiner_synthesis_2010,
	title = {Synthesis and Characterization of {K}$_{8-x}$({H}$_2$)$_y${Si}$_{46}$},
	volume = {49},
	issn = {0020-1669, 1520-510X},
	url = {https://pubs.acs.org/doi/10.1021/ic9004592},
	doi = {10.1021/ic9004592},
	pages = {815--822},
	number = {3},
	journal = {Inorganic Chemistry},
	shortjournal = {Inorg. Chem.},
	author = {Neiner, Doinita and Okamoto, Norihiko L. and Yu, Ping and Leonard, Sharon and Condron, Cathie L. and Toney, Michael F. and Ramasse, Quentin M. and Browning, Nigel D. and Kauzlarich, Susan M.},
	urldate = {2024-01-09},
	date = {2010-02-01},
    year = {2010},
	langid = {english},
}

@article{himeno_optical_2013,
  title = {Optical absorption properties of {Na}$_x${Si}$_{136}$ clathrate studied by diffuse reflection spectroscopy},
  author = {Himeno, Roto and Kume, Tetsuji and Ohashi, Fumitaka and Ban, Takayuki and Nonomura, Shuichi},
  journal = {Journal of Alloys and Compounds},
  volume = {574},
  pages = {398--401},
  year = {2013},
  doi = {10.1016/j.jallcom.2013.05.176},
  url = {https://linkinghub.elsevier.com/retrieve/pii/S0925838813013364}
}

@article{he_si-based_2014,
  author  = {He, Yuping and Sui, Fan and Kauzlarich, Susan M. and Galli, Giulia},
  title   = {Si-based Earth abundant clathrates for solar energy conversion},
  journal = {Energy Environ. Sci.},
  year    = {2014},
  volume  = {7},
  number  = {8},
  pages   = {2598--2602},
  doi     = {10.1039/C4EE00256C},
  url     = {http://xlink.rsc.org/?DOI=C4EE00256C}
}

@article{kawaji_superconductivity_1995,
  author  = {Kawaji, Hitoshi and Horie, Hiro-omi and Yamanaka, Shoji and Ishikawa, Mitsuo},
  title   = {Superconductivity in the Silicon Clathrate Compound ({Na},{Ba})$_x${Si}$_{46}$},
  journal = {Physical Review Letters},
  year    = {1995},
  volume  = {74},
  number  = {8},
  pages   = {1427--1429},
  doi     = {10.1103/PhysRevLett.74.1427},
  url     = {https://link.aps.org/doi/10.1103/PhysRevLett.74.1427}
}

@article{anderson_equation_1973,
	title = {Equation of state for solid neon to 20 kbar},
	volume = {10},
	rights = {http://www.springer.com/tdm},
	issn = {0022-2291, 1573-7357},
	url = {http://link.springer.com/10.1007/BF00654913},
	doi = {10.1007/BF00654913},
	pages = {345--357},
	number = {3},
	journal = {Journal of Low Temperature Physics},
	shortjournal = {J Low Temp Phys},
	author = {Anderson, M. S. and Fugate, R. Q. and Swenson, C. A.},
	urldate = {2024-10-21},
	date = {1973-02},
    year = {1973},
	langid = {english},
}

@article{song_straightforward_2021,
	title = {A straightforward approach to high purity sodium silicide {Na}$_{\textrm{4}}${Si}$_{\textrm{4}}$},
	volume = {50},
	issn = {1477-9226, 1477-9234},
	url = {https://xlink.rsc.org/?DOI=D1DT03203H},
	doi = {10.1039/D1DT03203H},
	pages = {16703--16710},
	number = {45},
	journal = {Dalton Transactions},
	shortjournal = {Dalton Trans.},
	author = {Song, Yang and Gómez-Recio, Isabel and Kumar, Ram and Coelho Diogo, Cristina and Casale, Sandra and Génois, Isabelle and Portehault, David},
	urldate = {2024-11-19},
	date = {2021},
    year = {2021},
	langid = {english},
}

@article{courac_high-pressure_2019,
	title = {High-Pressure Melting Curve of Zintl Sodium Silicide Na$_{\textrm{4}}$ Si$_{\textrm{4}}$ by In Situ Electrical Measurements},
	volume = {58},
	rights = {https://doi.org/10.15223/policy-029},
	issn = {0020-1669, 1520-510X},
	url = {https://pubs.acs.org/doi/10.1021/acs.inorgchem.9b01108},
	doi = {10.1021/acs.inorgchem.9b01108},
	pages = {10822--10828},
	number = {16},
	journal = {Inorganic Chemistry},
	shortjournal = {Inorg. Chem.},
	author = {Courac, Alexandre and Le Godec, Yann and Renero-Lecuna, Carlos and Moutaabbid, Hicham and Kumar, Ram and Coelho-Diogo, Cristina and Gervais, Christel and Portehault, David},
	urldate = {2024-11-19},
	date = {2019-08-19},
    year = {2019},
	langid = {english},
}

@article{debord_isostructural_2021,
	title = {Isostructural phase transition by point defect reorganization in the binary type-{I} clathrate {Ba}$_{7.5}${Si}$_{45}$},
	volume = {210},
	issn = {13596454},
	url = {https://linkinghub.elsevier.com/retrieve/pii/S1359645421002044},
	doi = {10.1016/j.actamat.2021.116824},
	pages = {116824},
	journal = {Acta Materialia},
	shortjournal = {Acta Materialia},
	author = {Debord, Régis and Euchner, Holger and Pischedda, Vittoria and Hanfland, Michael and San-Miguel, Alfonso and Mélinon, Patrice and Pailhès, Stéphane and Machon, Denis},
	urldate = {2024-11-26},
	date = {2021-05},
    year = {2021},
	langid = {english},
}

@article{jung_impact_2021,
	title = {The impact of boron atoms on clathrate-I silicides: composition range of the borosilicide {K}$_{\textrm{8-x}}${B}$_{\textrm{y}}${Si}$_{\textrm{46-y}}$},
	volume = {50},
	issn = {1477-9226, 1477-9234},
	url = {https://xlink.rsc.org/?DOI=D0DT03339A},
	doi = {10.1039/D0DT03339A},
	shorttitle = {The impact of boron atoms on clathrate-I silicides},
	pages = {1274--1282},
	number = {4},
	journal = {Dalton Transactions},
	shortjournal = {Dalton Trans.},
	author = {Jung, Walter and Böhme, Bodo and Hübner, Julia M. and Burkhardt, Ulrich and Borrmann, Horst and Bobnar, Matej and Nguyen, Hong Duong and Pantenburg, Ingo and Etter, Martin and Schwarz, Ulrich and Grin, Yuri and Baitinger, Michael},
	urldate = {2024-11-28},
	date = {2021},
    year = {2021},
	langid = {english},
}

@article{iitaka_pressure-induced_2007,
	title = {Pressure-induced isostructural phase transition of metal-doped silicon clathrates},
	volume = {75},
	rights = {http://link.aps.org/licenses/aps-default-license},
	issn = {1098-0121, 1550-235X},
	url = {https://link.aps.org/doi/10.1103/PhysRevB.75.012106},
	doi = {10.1103/PhysRevB.75.012106},
	pages = {012106},
	number = {1},
	journal = {Physical Review B},
	shortjournal = {Phys. Rev. B},
	author = {Iitaka, Toshiaki},
	urldate = {2024-12-05},
	date = {2007-01-16},
    year = {2007},
	langid = {english},
}

@article{sung_pressure_1977,
	title = {Pressure distribution in the diamond anvil press and the shear strenght of fayalite},
	volume = {48},
	issn = {0034-6748, 1089-7623},
	url = {https://pubs.aip.org/rsi/article/48/11/1386/308363/Pressure-distribution-in-the-diamond-anvil-press},
	doi = {10.1063/1.1134902},
	pages = {1386--1391},
	number = {11},
	journal = {Review of Scientific Instruments},
	author = {Sung, Chien-Min and Goetze, Christopher and Mao, Ho-Kwang},
	urldate = {2024-12-11},
	date = {1977-11-01},
    year = {1977},
	langid = {english},
}

@article{hubner_cage_2021,
	title = {Cage Adaption by High-Pressure Synthesis: The Clathrate-I Borosilicide {Rb}$_8${B}$_8${Si}$_{38}$},
	volume = {60},
	issn = {0020-1669},
	url = {https://doi.org/10.1021/acs.inorgchem.0c02357},
	doi = {10.1021/acs.inorgchem.0c02357},
	shorttitle = {Cage Adaption by High-Pressure Synthesis},
	pages = {2160--2167},
	number = {4},
	journal = {Inorganic Chemistry},
	shortjournal = {Inorg. Chem.},
	author = {Hübner, Julia-Maria and Jung, Walter and Schmidt, Marcus and Bobnar, Matej and Koželj, Primož and Böhme, Bodo and Baitinger, Michael and Etter, Martin and Grin, Yuri and Schwarz, Ulrich},
	urldate = {2024-12-12},
	date = {2021-02-15},
    year = {2021},
	note = {Publisher: American Chemical Society},
}

@article{jung_k7_2007,
	title = {{K}$_{\textrm{7}}${B}$_{\textrm{7}}${Si}$_{\textrm{39}}$ , a Borosilicide with the Clathrate I Structure},
	volume = {46},
	rights = {http://onlinelibrary.wiley.com/{termsAndConditions}\#vor},
	issn = {1433-7851, 1521-3773},
	url = {https://onlinelibrary.wiley.com/doi/10.1002/anie.200701028},
	doi = {10.1002/anie.200701028},
	pages = {6725--6728},
	number = {35},
	journal = {Angewandte Chemie International Edition},
	shortjournal = {Angew Chem Int Ed},
	author = {Jung, Walter and Lörincz, Josef and Ramlau, Reiner and Borrmann, Horst and Prots, Yurii and Haarmann, Frank and Schnelle, Walter and Burkhardt, Ulrich and Baitinger, Michael and Grin, Yuri},
	urldate = {2024-12-13},
	date = {2007-09-03},
    year = {2007},
	langid = {english},
}

@article{gonzalez-platas_eosfit7-gui_2016,
	title = {\textit{{EosFit}7-{GUI}} : a new graphical user interface for equation of state calculations, analyses and teaching},
	volume = {49},
	issn = {1600-5767},
	url = {https://journals.iucr.org/paper?S1600576716008050},
	doi = {10.1107/S1600576716008050},
	shorttitle = {\textit{{EosFit}7-{GUI}}},
	pages = {1377--1382},
	number = {4},
	journal = {Journal of Applied Crystallography},
	shortjournal = {J Appl Crystallogr},
	author = {Gonzalez-Platas, Javier and Alvaro, Matteo and Nestola, Fabrizio and Angel, Ross},
	urldate = {2025-04-16},
	date = {2016-08-01},
    year = {2016},
	langid = {english},
}

@article{gerin_structural_2023,
	title = {Structural transitions at high pressure and metastable phase in {Si}$_{0.8}${Ge}$_{0.2}$},
	volume = {954},
	issn = {09258388},
	url = {https://linkinghub.elsevier.com/retrieve/pii/S0925838823014834},
	doi = {10.1016/j.jallcom.2023.170180},
	pages = {170180},
	journal = {Journal of Alloys and Compounds},
	shortjournal = {Journal of Alloys and Compounds},
	author = {Gerin, M. and Machon, D. and Radescu, S. and Le Floch, S. and Le Godec, Y. and Gaudisson, T. and Alabarse, F. and Veber, P. and Debord, R. and Amans, D. and Pischedda, V.},
	urldate = {2025-04-16},
	date = {2023-09},
    year = {2023},
	langid = {english},
}

@article{zhang_covalent_2018,
	title = {Covalent Radii and New Applications:},
	volume = {7},
	issn = {2155-4110, 2155-4129},
	url = {http://services.igi-global.com/resolvedoi/resolve.aspx?doi=10.4018/IJCCE.2018010103},
	doi = {10.4018/IJCCE.2018010103},
	shorttitle = {Covalent Radii and New Applications},
	pages = {42--51},
	number = {1},
	journal = {International Journal of Chemoinformatics and Chemical Engineering},
	author = {Zhang, Yonghe},
	urldate = {2025-04-17},
	date = {2018-01},
    year = {2018},
	langid = {english},
}

@article{loubeyre_synchrotron_2020,
	title = {Synchrotron infrared spectroscopic evidence of the probable transition to metal hydrogen},
	volume = {577},
	issn = {0028-0836, 1476-4687},
	url = {https://www.nature.com/articles/s41586-019-1927-3},
	doi = {10.1038/s41586-019-1927-3},
	pages = {631--635},
	number = {7792},
	journal = {Nature},
	shortjournal = {Nature},
	author = {Loubeyre, Paul and Occelli, Florent and Dumas, Paul},
	urldate = {2025-04-19},
	date = {2020-01-30},
    year = {2020},
	langid = {english},
}

@article{vegard_konstitution_1921,
	title = {Die Konstitution der Mischkristalle und die Raumfüllung der Atome},
	volume = {5},
	issn = {0044-3328},
	url = {https://doi.org/10.1007/BF01349680},
	doi = {10.1007/BF01349680},
	pages = {17--26},
	number = {1},
	journal = {Zeitschrift für Physik},
	shortjournal = {Z. Physik},
	author = {Vegard, L.},
	urldate = {2025-05-09},
	date = {1921-01-01},
    year = {1921},
	langid = {german},
}

@article{san-miguel_pressure_2002,
    title = {Pressure stability and low compressibility of intercalated cagelike materials: The case of silicon clathrates},
    volume = {65},
    copyright = {http://link.aps.org/licenses/aps-default-license},
    issn = {0163-1829, 1095-3795},
    shorttitle = {Pressure stability and low compressibility of intercalated cagelike materials},
    url = {https://link.aps.org/doi/10.1103/PhysRevB.65.054109},
    doi = {10.1103/PhysRevB.65.054109},
    number = {5},
    urldate = {2024-12-05},
    journal = {Physical Review B},
    author = {San-Miguel, A. and Mélinon, P. and Connétable, D. and Blase, X. and Tournus, F. and Reny, E. and Yamanaka, S. and Itié, J. P.},
    month = jan,
    year = {2002},
    pages = {054109},
}

@article{miguel_pressure-induced_2005,
    title = {Pressure-induced homothetic volume collapse in silicon clathrates},
    volume = {69},
    issn = {0295-5075, 1286-4854},
    url = {https://iopscience.iop.org/article/10.1209/epl/i2004-10387-x},
    doi = {10.1209/epl/i2004-10387-x},
    number = {4},
    urldate = {2024-12-05},
    journal = {Europhysics Letters (EPL)},
    author = {Miguel, A. San and Merlen, A and Toulemonde, P and Kume, T and Floch, S. Le and Aouizerat, A and Pascarelli, S and Aquilanti, G and Mathon, O and Bihan, T. Le and Itié, J.-P and Yamanaka, S},
    month = feb,
    year = {2005},
    pages = {556--562},
}

\end{document}